\documentclass[aps,prb,twocolumn,showpacs,floatfix,groupedaddress
] {revtex4-1}
\usepackage[T1]{fontenc}
\usepackage[latin9]{inputenc}

\usepackage{graphicx}
\usepackage{amssymb}
\usepackage{amsmath}
\usepackage{color}



\begin{document}

\newcommand{\be}{\begin{equation}}
\newcommand{\ee}{  \end{equation}}
\newcommand{\ba}{\begin{eqnarray}}
\newcommand{\ea}{  \end{eqnarray}}
\newcommand{\ve}{\varepsilon}

\title{Photon-assisted electronic and spin transport in a junction containing precessing molecular spin}

\author{Milena Filipovi\'{c}}
\author{Wolfgang Belzig}
\affiliation{Fachbereich Physik, Universit\"at Konstanz, D-78457 Konstanz, Germany}

\date{\today}

\begin{abstract}
We study the ac charge and -spin transport through an orbital of a magnetic molecule with spin precessing in a constant magnetic field. We assume that the source and drain contacts have time-dependent chemical potentials. We employ the Keldysh nonequilibrium Green's functions method to calculate the spin and charge currents to linear order in the time-dependent potentials. The molecular and electronic spins are coupled via exchange
interaction. The time-dependent molecular spin drives inelastic transitions between the molecular quasienergy levels, resulting in a rich structure in the transport characteristics. The time-dependent voltages allow us to reveal the internal precession time scale (the Larmor frequency) by a dc conductance measurement if the ac frequency matches the Larmor frequency. In the low-ac-frequency limit the junction resembles a classical electric circuit. Furthermore, we show that the setup can be used to generate dc-spin currents, which are controlled by the molecular magnetization direction and the relative phases between the Larmor precession and the ac voltage.
\end{abstract}
\pacs{73.23.-b, 75.76.+j, 85.65.+h, 85.75.-d}
\maketitle

\section{Introduction}

Since the early 1970s, the potential use of molecules as components of electronic circuitry was proposed,\cite{Ratner1} thereby introducing the field of molecular electronics.  
Since then, the goal of the field has been to 
create  high-speed processing molecular devices with miniature size.\cite{Ratner2,Elke1}
In that respect, it is important to investigate the properties of transport through single molecules in the presence of 
external fields.\cite{Peter,Dulic,Molen,KimElke,Elke2}
Single-molecule magnets are a class of molecular magnets with a large spin, strong magnetic anisotropy, and slow magnetization relaxation at low temperatures.\cite{ses} Due to both classical\cite{hysteresis} and 
quantum\cite{hysteresis,spintunneling,coherence,interference} 
characteristics of single-molecule magnets, their application in molecular electronics became a topic of intense 
research, considering their potential usage in creation of memory devices.\cite{Flat} Several experiments have already achieved transport through single-molecule magnets.\cite{Voss14,Burzuri14,Kahle14}

Time-dependent transport through molecular junctions has been theoretically studied using 
different techniques, such as nonequilibrium Green's functions technique,\cite{Jauho1993,Jauho1994,JauhoBook,Dong,Zhu} time-dependent density functional 
theory,\cite{dft1,dft2,dft3,dft4,dft5} reduced density matrix approach,\cite{Welack} etc. 
Time-dependent periodic fields in electrical contacts 
cause photon-assisted tunneling,\cite{Martin,Gordon,PlateroAguado, Peter} a phenomenon based on the fact that by applying an external 
harmonic field with frequency $\Omega$ to the contact, the conduction
electrons interact with the ac field and, consequently, participate in the inelastic tunneling processes 
by absorbing or emitting an amount of energy $n\hbar\Omega$, where $n=\pm 1,\pm 2,....$ Theoretically, photon-assisted tunneling through atoms and molecules
was investigated in numerous works.\cite{Peter, Zheng01, Tikhonov, Schreiber, Vilj, Hsu, Fainberg} 
Some experimental studies addressed photon-assisted tunneling through atomic-sized\cite{Chauvin,Guhr,Noy1} and molecular\cite{Noy2,Mangold} 
junctions in the presence of laser fields. 
Time-dependent electric control of the state of quantum spins 
of atoms has also been investigated.\cite{LothBergmann}
In junctions with time-dependent ac bias, the presence of displacement currents is inevitable 
due to the charge accumulation in the scattering region.\cite{But01,But02} This problem can be solved either implicitly 
by including the Coulomb interaction in the Hamiltonian of the system\cite{But03,Wei01} or explicitly by adding the displacement 
current to the conduction current,\cite{But02,Guo1} thus providing the conservation of the total ac current. 

Spin transport through magnetic nanostructures can be used to manipulate the state of the magnetization via spin-transfer torques (STTs).\cite{slonc,berg} The concept of STT is based on the transfer of spin angular momenta from the conduction electrons to a local magnetization in the scattering region, generating a torque as a back-action of the spin transport, and thus changing the state of the magnetic nanostructure.\cite{slonc,berg,tser,ralp} Hence, current-induced magnetization reversal became an active topic in recent years.\cite{Slon1,Timm,Misiorny,Slon2,Xiao1,Delgado,Bode} The measurement and control of the magnetization of single-molecule magnets employing spin transport may bring important applications in spintronics. 

In this work we theoretically study the charge and spin transport through
a single electronic energy level in the presence of a molecular spin in a constant magnetic field. The  electronic level may be an orbital of the molecule or it may belong to a nearby quantum dot. The molecular spin, treated as a classical magnetic moment, exhibits Larmor precession around the magnetic field axis. The Zeeman field and interaction of the orbital with the precessing molecular spin result in four quasienergy levels in the quantum dot, obtained using the Floquet theorem.\cite{Floquet1,Floquet2,Floquet3,Floquet4} The  system is then connected to electric contacts subject to oscillating electric potentials, considered as a perturbation. 
The oscillating chemical potentials induce photon-assisted charge and spin tunneling.  
A photon-assisted STT is exerted on the molecular spin by the photon-assisted spin-currents. This torque is not included in the dynamics of the molecular spin, since the molecular spin precession is assumed to be kept steady by external means, thus compensating the STT. The precessing molecular spin in turn pumps spin-currents into the leads, acting as an external rotating exchange field.
Some of our main results are as follows: 
\begin{enumerate}
	\item In the limit of low ac frequency, the junction can be mapped onto a classical electric circuit modeling the inductive-like or capacitive-like response.
	\item The real and imaginary components of the dynamic conductance, associated with the resonant position of the chemical potentials with molecular quasienergy levels, are both enhanced around the ac frequency matching the Larmor frequency, allowing the detection of the internal precession time scale (see Fig.~\ref{fig: frequency}).
	\item The setup can be employed to generate and control dc spin currents by tuning the molecular precession angle and the relative phases between the ac voltage and Larmor precession if ac frequency matches the Larmor frequency.
\end{enumerate} 

A part of this article is a complement to Ref.\citenum{we}, representing the solution for the Gilbert damping coefficient,\cite{Gilbert} nonperturbative in the coupling to the molecular magnet, in the absence of time-varying voltage. The other corresponding STT coefficients and an arising nonzero $z$ component of the STT are obtained as well.

The article is organized in the following way: We describe the model setup of the system in Sec.~II. The theoretical formalism based on the Keldysh nonequilibrium Green's functions technique\cite{Jauho1993,Jauho1994,JauhoBook} is introduced in Sec.~III. Here we derive expressions for spin and charge currents in linear order with respect to ac harmonic potentials in the leads. In Sec.~IV we obtain and analyze the dynamic conductance of the charge current using the current partitioning scheme developed by Wang \textit{et al}.\cite{Guo1} This section is followed by Sec.~V in which we analyze spin transport and STT under dc-bias voltage and in the presence of oscillating chemical potentials. We finally conclude in Sec.~VI.	

\section{Model setup}
\begin{figure}
\includegraphics[width=8.5cm,keepaspectratio=true]{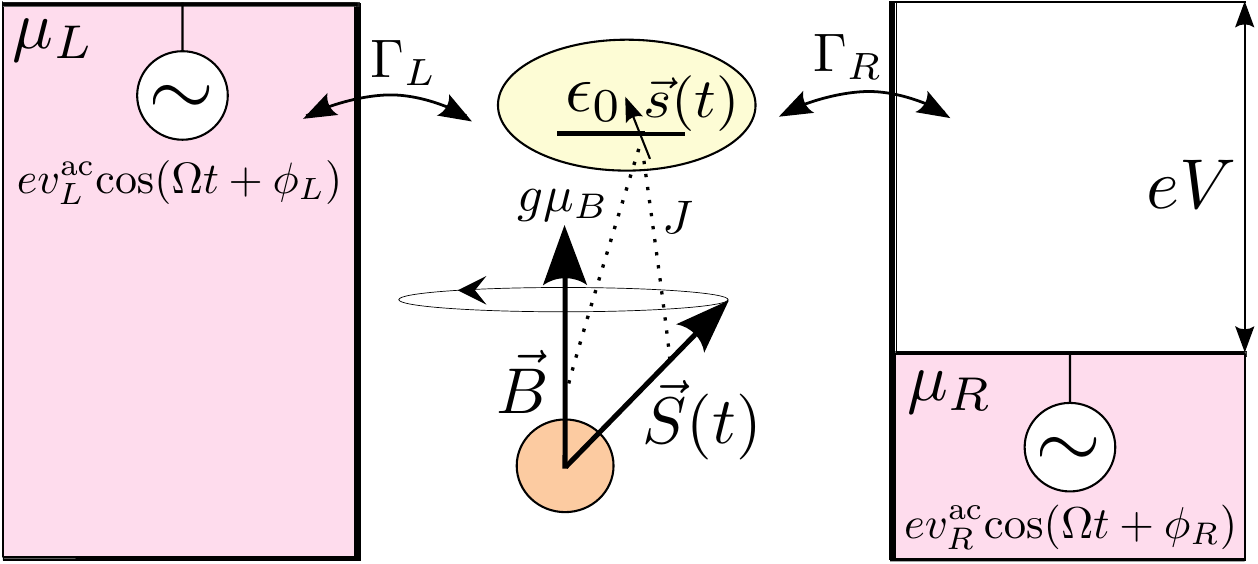}
\caption{(Color online) Photon-assisted tunneling through 
	a single molecular level with energy $\epsilon_{0}$ coupled to the spin $\vec{S}(t)$ of a molecular magnet via  exchange interaction
	with the coupling constant $J$,
	in the presence of a constant 
	magnetic field $\vec{B}$.~External ac potentials
	$V^\textrm{ac}_{\xi}(t)=v^\textrm{ac}_{\xi}{\rm cos}(\Omega t+\phi_\xi)$ are applied to the leads $\xi=L,R$ with chemical potentials $\mu_{\xi}$
	and tunnel rates $\Gamma_{\xi}$.}\label{fig: system}
\end{figure}
We consider a junction consisting of a single spin-degenerate molecular orbital of a molecular magnet with a precessing spin in a constant magnetic field along $z$-axis, $\vec{B}=B\vec{e}_{z}$, coupled to two normal metallic leads. 
We assume the spin of the molecular magnet is large and neglecting the quantum fluctuations treat it as a classical 
vector $\vec{S}$, with constant length $S=\textbar{\vec{S}}\textbar$. 
The magnetic field does not affect the electric contacts, which are assumed to be noninteracting. An external ac harmonic potential 
$V_{\xi}^\mathrm{ac}(t)=v_{\xi}^\mathrm{ac}{\mathrm{cos}}(\Omega t+\phi_\xi)$ is applied to each lead $\xi=L,R$, modulating the single electron energy as $\epsilon_{k\xi }(t)=\epsilon_{k\xi}+eV_{\xi}^\mathrm{ac}(t)$, 
with $\epsilon_{k\xi}$ being the single-particle energy of an electron with the wave number $k$, 
in the absence of the time-varying voltage (see Fig.~\ref{fig: system}). Since we want to unravel the quantum effects induced by the tunneling electrons and the ac harmonic potentials, we consider a well coupled molecular orbital and treat it as noninteracting by disregarding the intraorbital Coulomb interactions between the electrons.
 
The junction is described by the Hamiltonian $\hat{H}(t)=\hat{H}_{L}(t)+\hat{H}_{R}(t)+\hat{H}_{T}+\hat{H}_{MO}(t)+\hat{H}_{S}$. 
Here $\hat{H}_\xi(t)=\sum_{k,\sigma}\epsilon_{k\xi}(t) \hat{c}^\dagger_{k\sigma\xi} \hat{c}_{k\sigma\xi}$ 
is the Hamiltonian of lead $\xi=L,R$. The subscript $\sigma=\uparrow,\downarrow=1,2=\pm 1$ denotes the spin-up or spin-down state of the electrons. The tunneling Hamiltonian $\hat{H}_{T}=\sum_{k,\sigma,\xi}  [V_{k\xi}\hat{c}^\dagger_{k\sigma\xi} \hat{d}_{\sigma}+V^{\ast}_{k\xi}
\hat{d}^\dagger_{\sigma} \hat{c}_{k\sigma\xi}]$ introduces the spin-independent tunnel coupling between the molecular orbital and the leads, with matrix element $V_{k\xi}$. The 
operators $ \hat{c}^\dagger_{k\sigma\xi}(\hat{c}_{k\sigma\xi})$ and $ \hat{d}^\dagger_{\sigma} (\hat{d}_{\sigma})$ represent 
the creation (annihilation) operators of the electrons in the leads and the molecular orbital. The next term in the Hamiltonian of the system is given 
by $\hat{H}_{MO}(t)=\sum_{{\sigma}}\epsilon_{0} \hat{d}^\dagger_{\sigma} \hat{d}_{\sigma}+(g\mu_{B}/\hbar)\hat{\vec{s}}\vec B+J\hat{\vec{s}}\vec{S}(t)$. 
Here, the first term describes the noninteracting molecular orbital with energy $\epsilon_{0}$. The second term represents the electronic spin in the molecular orbital, $\hat{\vec{s}}=(\hbar/2)\sum_{\sigma\sigma'}(\hat{\vec\sigma})_{\sigma\sigma'}\hat d^\dagger_\sigma\hat d_{\sigma'}$, in the presence of the external constant magnetic field $\vec{B}$, and the third term expresses the exchange interaction between the electronic spin and the molecular spin $\vec{S}(t)$. Here, $\hat{\vec\sigma}=(\hat{\sigma}_x,\hat{\sigma}_y,\hat{\sigma}_z)^T$ represents the vector of the Pauli matrices.
The proportionality factors $g$ and $\mu_{B}$ are the gyromagnetic ratio of the electron and the Bohr magneton, respectively,  while $J$ is the exchange coupling constant between the molecular and electronic spins. 

Presuming for simplicity, that the molecular spin $g$ factor equals that of a free electron, the term $\hat{H}_{S}=g\mu_{B}{\vec S} {\vec B}$ represents the energy of
the classical molecular spin $\vec S$ in the magnetic field $\vec B$.
Accordingly, the field $\vec B$ exerts a torque on the spin $\vec S$ leading to its precession around the field axis with Larmor frequency $\omega_L=g\mu_{B}B/\hbar$.
To compensate for the dissipation of magnetic energy due to the interaction with conduction electrons, we assume that the molecular spin is kept precessing by external means (e.g., rf fields).\cite{Kittel} Hence, we keep the tilt angle $\theta$ between $\vec{B}$ and $\vec{S}$ fixed and determined by the initial conditions. The dynamics of the molecular spin is then given by $\vec S(t)=S_{\bot}\cos (\omega_{L} t)\vec e_x+S_{\bot}\sin (\omega_{L} t)\vec e_y+S_{z}\vec e_z$, where $S_{\bot}$ is the magnitude of the instantaneous projection of $\vec S(t)$ onto the $x$-$y$ plane, given by $S_{\bot}=S\sin(\theta)$, while the projection of the molecular
spin on the $z$-axis equals $S_z=S\cos(\theta)$. The precessing spin $\vec S(t)$ pumps spin-currents into the system, but the effects of spin currents onto the molecular spin dynamics are compensated by the above-mentioned external sources. 

\section{Theoretical Formalism}

The ensemble and quantum average charge and spin currents from the lead $\xi$ to the molecular orbital are given by 
\begin{equation}
I_{\xi\nu}(t) =q_\nu\bigg \langle \frac{d}{dt} \hat{N}_{\xi\nu} \bigg \rangle\ =  q_\nu\frac{i}{\hbar} 
\big \langle \big [\hat{H},\hat{N}_{\xi\nu} \big ] \big \rangle,
\end{equation}
with $\hat{N}_{\xi\nu}=\sum_{{k,\sigma,\sigma\prime}}
\hat{c}^\dagger_{k\sigma\xi}({{\sigma}}_\nu)_{\sigma\sigma^\prime} 
\hat{c}_{k\sigma\prime\xi}$ representing the charge and spin occupation number operator of the contact $\xi$. 
The index $\nu$ takes values $\nu=0$ for the charge and $\nu= 1,2,3$ for the components $x,y,z$ 
of the spin-polarized current. The prefactors $q_\nu$ 
correspond to the electronic charge $q_0=-e$ and spin $q_{\nu\neq 0}=\hbar/2$.
Employing the Keldysh nonequilibrium Green's functions technique, the currents can be calculated in units in which $\hbar=e=1$ as \cite{Jauho1994, JauhoBook}
\begin{align}
\label{eq: general current}
I_{\xi\nu}{(t)} =&-{2q_\nu}{\rm Re}\int dt^\prime {\rm Tr}\big\{\hat\sigma_\nu[ {\hat{G}}^{r} {(t,t^\prime)}{\hat{\Sigma}}^{<}_{\xi}(t^\prime,t)\nonumber\\
& \quad \quad \quad\quad \quad\quad +{\hat{G}}^{<} {(t,t^\prime)}{\hat{\Sigma}}^{a}_{\xi}(t^\prime,t) ]\big\},
\end{align}
where ${\hat\sigma_0}=\hat 1$ is the identity operator, while $\hat\sigma_{\nu\neq 0}$ are the Pauli matrices.
In Eq.~(\ref{eq: general current}), $\hat{\Sigma}^{r,a,<}_{\xi}(t,t^\prime)$ 
are the retarded, advanced, and lesser self-energies from the tunnel coupling between the molecular orbital
and the lead $\xi$, while $\hat{G}^{r,a,<}(t,t^\prime)$ are the corresponding Green's functions of the electrons in the molecular orbital.
The matrices of the self-energies are diagonal in the electronic spin space with respect to the basis of eigenstates 
of $\hat{s}_{z}$, and their nonzero entries are given by $\Sigma^{r,a,<}_{\xi}(t,t')=\sum_{{k}}
V_{k\xi}^{\phantom{\ast}}g^{r,a,<}_{k\xi}{(t,t')}V^{\ast}_{k\xi}$,
where $g^{r,a,<}_{k \xi}{(t,t')}$ are the retarded, advanced and lesser Green's functions of the electrons in contact $\xi$.
The matrix elements of the Green's functions $\hat{G}^{r,a,<}(t,t^\prime)$ are given by 
$G^{r,a}_{\sigma\sigma^\prime} {(t,t^\prime)}=\mp i\theta(\pm t \mp t^\prime)\langle\{\hat{d}_{\sigma}{(t)},
\hat{d}^\dagger_{\sigma^\prime} {(t^\prime)}\}\rangle$ and $G^<_{\sigma\sigma^\prime} (t,t^\prime)= i \langle \hat{d}^\dagger_{\sigma^\prime} (t^\prime) \hat{d}_\sigma(t)\rangle$, where $\{\cdot ,\cdot\}$ denotes the anticommutator.
The self-energies of lead $\xi$ can be expressed as\cite{Jauho1993,Jauho1994,JauhoBook}
\begin{align}
	{\Sigma}^{<}_{\xi}(t,t^\prime) & =
	i\int\frac{d\epsilon}{2\pi}e^{-i\epsilon (t-t^\prime)+i\varphi_\xi(t,t')}
	f_{\xi}(\epsilon)\Gamma_{\xi}(\epsilon),\label{eq: selff}\\
	{\Sigma}^{r}_{\xi}(t,t^\prime) & =
	-i\theta(t-t^\prime)\int\frac{d\epsilon}{2\pi}
	e^{-i\epsilon (t-t^\prime)+i\varphi_\xi(t,t')}
	\Gamma_{\xi}(\epsilon).
\end{align}
Here we introduced the Faraday phases $\varphi_\xi(t,t')=e\int^{t^\prime}_{t}dt^{\prime\prime}V^{\rm{ac}}_{\xi}( t^{\prime\prime})$. From its definition, it follows that ${\Sigma}^{a}_{\xi}(t,t^\prime)= [{\Sigma}^{r}_{\xi}(t^{\prime},t)]^*$. Furthermore, $f_{\xi}(\epsilon)=[e^{(\epsilon-\mu_{\xi})/k_{B}T}+1]^{-1}$ is the Fermi-Dirac distribution of the electrons in the lead $\xi$,
with $k_{B}$ the Boltzmann constant and $T$ the temperature, while $\Gamma_{\xi}(\epsilon)=2\pi\sum_{{k}}\lvert V_{k\xi}\rvert^{2}\delta (\epsilon-\epsilon_{k\xi})$ is the tunnel coupling to the lead $\xi$. 
Using the self-energies defined above, and applying the double Fourier transformations in Eq.~(\ref{eq: general current}), in the wide-band limit, in which $\Gamma_{\xi}$ is energy independent, one obtains 
\begin{align}\label{eq: Bessel}
I_{\xi\nu}{(t)}=&\,2q_{\nu}\Gamma_{\xi}{\rm Im}\int\frac{d\epsilon}{2\pi}\int\frac{d\epsilon^\prime}{2\pi} e^{-i(\epsilon-\epsilon^\prime)t}\nonumber\\&\times\displaystyle\sum\limits_{m,n}{J_{m}\left(\frac{v^{\rm{ac}}_{\xi}}{\Omega}\right)}{J_{n}\left(\frac{v^{\rm{ac}}_{\xi}}{\Omega}\right)}e^{i(m-n)\phi_{\xi}}\nonumber\\
&\times{\rm Tr}\bigg\{\hat\sigma_\nu\bigg [f_{\xi}(\epsilon^{\prime}_{m}){\hat{G}}^{r} {(\epsilon,\epsilon^{\prime}_{mn})+ {\frac{1}{2}{\hat{G}}^{<}} {(\epsilon,\epsilon^{\prime}_{mn})}\bigg ]}\bigg\},
\end{align}
with the abbreviations
$\epsilon_{m}=\epsilon-m\Omega$ and
$\epsilon_{mn}=\epsilon-(m-n)\Omega$. The generating function $\exp[ia\sin(\Omega t+\phi)]=\sum_m J_{m}(a)\exp[im(\Omega t +\phi)]$ was used in Eq.~(\ref{eq: Bessel}), where $J_{m}$ is the Bessel function of the first kind of order $m$.

The matrix components of the retarded Green's function of the electrons in the molecular orbital, in the absence of the ac harmonic potentials in the leads, 
can be obtained exactly by applying Dyson's expansion and analytic continuation rules.\cite{JauhoBook} Their double Fourier transforms are written as\cite{Guo}
\begin{align}
	\mathcal{G}^{r}_{\sigma\sigma}(\epsilon,\epsilon^\prime)&
	=\frac{2\pi \delta(\epsilon-\epsilon^\prime)G^{0r}_{\sigma\sigma}(\epsilon)}{
		1-\gamma^{2}G^{0r}_{\sigma\sigma}(\epsilon)
			{G^{0r}_{-\sigma-\sigma}(\epsilon_{\sigma})}},\\
	\mathcal{G}^{r}_{\sigma-\sigma}(\epsilon,\epsilon^\prime)&
	=\frac{2\pi\gamma\delta(\epsilon_{\sigma}-\epsilon^{\prime})
	G^{0r}_{\sigma\sigma}(\epsilon)
	G^{0r}_{-\sigma-\sigma}(\epsilon_{\sigma})}{
		1-\gamma^{2}G^{0r}_{\sigma\sigma}(\epsilon)
		G^{0r}_{-\sigma-\sigma}(\epsilon_{\sigma})},
\end{align}
with $\gamma=JS\sin(\theta)/2$ and $\epsilon_{\sigma}=\epsilon-\sigma\omega_L$. The matrix elements of the 
corresponding lesser Green's function are obtained using the Fourier 
transfomed Keldysh equation $\hat{\mathcal{G}}^{<}(\epsilon,\epsilon^\prime)=\int d\epsilon^{\prime\prime}\hat{\mathcal{G}}^{r}(\epsilon,\epsilon^{\prime\prime})\hat{\Sigma}^{<}_{0}(\epsilon^{\prime\prime})\hat{\mathcal{G}}^{a}(\epsilon^{\prime\prime},\epsilon^\prime)/2\pi$.\cite{JauhoBook} Here $\hat{\mathcal{G}}^{a}(\epsilon,\epsilon^\prime)=[\hat{\mathcal{G}}^{r}(\epsilon^\prime,\epsilon)]^\dagger$ and ${\Sigma}^{<}_{0}(\epsilon)=i\sum_{\xi}\Gamma_{\xi}f_{\xi}(\epsilon)$ is the lesser self-energy originating from the orbital-lead coupling in the absence of harmonic potentials in the leads. 
The retarded Green's functions $\hat{G}^{0r}$ of the electrons in the molecular orbital, in
the presence of the static component of the molecular spin and the constant magnetic field $\vec B$, are found using the equation of motion technique\cite{Bruus} and, Fourier transformed, read 
$\hat{G}^{0r}(\epsilon)=[\epsilon-\epsilon_{0}-\Sigma_{0}^{r}(\epsilon)-\hat{\sigma}_{z} (g\mu_{B}B+J S_{z})/2]^{-1}$,\cite{Guo,Bode} where $\Sigma^{r}_{0}(\epsilon)=-i\Gamma/2$ and $\Gamma=\sum_{\xi}\Gamma_{\xi}$.

For a weak ac field $v^{\rm{ac}}_{\xi}\ll\Omega$, the retarded and lesser Green's functions of the electrons in the molecular orbital can be 
obtained by applying Dyson's expansion, analytic continuation rules, and the Keldysh equation.\cite{JauhoBook} Keeping only terms linear in $v_{\xi}^\mathrm{ac}/\Omega$ they read
\begin{align}
	\hat{G}^{r}(\epsilon,\epsilon^\prime)\label{eq: final Green1} 
	\approx &\,\hat{\mathcal{G}}^{r}(\epsilon,\epsilon^\prime),\\
	\hat{G}^{<}(\epsilon,\epsilon^\prime )\label{eq: final Green2}
	\approx &\,\hat{\mathcal{G}}^{<}(\epsilon,\epsilon^\prime)+i
	\sum\limits_{\xi,n=\pm 1}n
	\Gamma_{\xi}\frac{v_{\xi}^\mathrm{ac}}{\Omega}e^{in\phi_{\xi}}
	\nonumber\\
	&\times \int\frac{d\epsilon''}{4\pi}[f_{\xi}(\epsilon''_{n})-f_{\xi}(\epsilon'')]
	\hat{\mathcal{G}}^{r}(\epsilon,\epsilon''_{n})\hat{\mathcal{G}}^{a}(\epsilon'',\epsilon').
\end{align}
In the rest of the paper we will stay in this limit.

The particle current contains the following contributions: 
\begin{equation}
	I_{\xi\nu}{(t)}=I^{\omega_{L}}_{\xi\nu}{(t)}+I^{\Omega}_{\xi\nu}{(t)}.
	\label{eq:particle_current}
\end{equation}
The first component represents the transport in the absence of ac voltages in the leads. It has a static and a time-dependent contribution, which are both created by 
the precession of the molecular spin. This precession-induced current reads
\begin{align}
	\label{eq:prec_ind_current}
	I_{\xi\nu}^{\omega_{L}}{(t)}
	=&\,2q_{\nu}\Gamma_{\xi}{\rm Im}\bigg\{\int\frac{d\epsilon}{2\pi}\int\frac{d\epsilon^\prime}{2\pi} 
	e^{-i(\epsilon-\epsilon^\prime)t}\nonumber\\
	&\times{\rm Tr}\bigg\{\hat\sigma_\nu\bigg [\frac{1}{2}\hat{\mathcal{G}}^{<}(\epsilon,\epsilon')+f_{\xi}(\epsilon')\hat{\mathcal{G}}^{r}(\epsilon,\epsilon')\bigg ]\bigg\}\bigg\}.
\end{align}
In the limit $\gamma^{2}\rightarrow 0$, Eq. (\ref{eq:prec_ind_current}) reduces to the result obtained previously.\cite{we}
The second term of Eq.~(\ref{eq:particle_current}) is induced when an ac voltage is applied to lead $\xi$ and can be expressed in linear order with respect to $v_{\xi}^\mathrm{ac}/\Omega$ using Eqs. (\ref{eq: Bessel}), (\ref{eq: final Green1}), and (\ref{eq: final Green2}) as
\begin{align}
	\label{eq:ac_voltage_current}
I_{\xi\nu}^{\Omega}{(t)}&=q_{\nu}\displaystyle\sum\limits_{\zeta,n=\pm 1}n\Gamma_{\xi}\Gamma_{\zeta}\frac{v_{\zeta}^{\rm {ac}}}{\Omega}{\rm Re}\int\frac{d\epsilon}{2\pi}\int\frac{d\epsilon^\prime}{2\pi} e^{-i(\epsilon-\epsilon^\prime)t+in\phi_{\zeta}}\nonumber\\
\times&\bigg\{\int\frac{d\epsilon''}{4\pi}\{[f_{\zeta}(\epsilon''_{n})-f_{\zeta}(\epsilon'')]{\rm Tr}[\hat\sigma_\nu\hat{\mathcal{G}}^{r}(\epsilon,\epsilon''_n)\hat{\mathcal{G}}^{a}(\epsilon'',\epsilon')]\}\nonumber\\
&-\frac{i}{\Gamma_{\zeta}}\delta_{\xi\zeta}[f_{\zeta}(\epsilon'_{n})-f_{\zeta}(\epsilon')]{\rm Tr}[\hat\sigma_\nu\hat{\mathcal{G}}^{r}(\epsilon,\epsilon'_n)]\bigg\}.
\end{align}
These expressions for the currents constitute the main results of the article. They allow us to calculate the dynamic charge conductance and spin transport properties of our molecular contact. Note that spin currents are more conveniently discussed in terms of the spin-transfer torque exerted by the inelastic spin currents onto the spin of the molecule, given by \cite{slonc,berg,tser,ralp}
\begin{equation}
	\label{eq: torque}
	\vec T(t)=\vec{T}^{\omega_L}(t)+\vec{T}^{\Omega}(t)=-[{\vec I}_L(t)+{\vec I}_R(t)]\,.
 \end{equation}
 Hence, in the remainder of the article we will concentrate on the ac charge conductance and the dc spin-transfer torque.

\section{Charge transport}
\subsection{Dynamic charge conductance}

The time-dependent particle charge current from the lead $\xi$ to the molecular orbital is induced by the ac harmonic potentials in the leads and can be written as
\begin{equation}
\label{eq:current_vac1}
	I^{\Omega}_{\xi 0}(t)=
	{\rm Re}\Bigg\{\displaystyle\sum\limits_{\zeta}G^{c}_{\xi\zeta}(\Omega)v_{\zeta}^{\rm{ac}}
	e^{-i(\Omega t+\phi_{\zeta})}\Bigg\},
\end{equation}
where $G^{c}_{\xi\zeta}(\Omega)$ is the conductance between leads $\xi$ and $\zeta$. 

In order to determine the dynamic conductance under ac bias-voltage conditions, one also needs to take into account the contribution 
from the displacement current. Coulomb interaction leads to screening of the charge accumulation in the quantum dot given by
$I^{d}(t)=\frac{dQ(t)}{dt}=-e{\rm Im}\{\frac{d}{dt}[{\rm Tr}\hat{G}^{<}(t,t)]\}$. According to the Kirchhoff's current law, $I^{d}(t)+\sum_{\xi}I^{\Omega}_{\xi 0}(t)=0$.
The following expression defines 
the total conductance of charge current, $G_{\xi\zeta}$:
\begin{equation}
I^{\Omega , tot}_{\xi 0}(t)={\rm Re}\Bigg\{\displaystyle\sum\limits_{\zeta}G_{\xi\zeta}(\Omega)v_{\zeta}^{\rm{ac}}e^{-i(\Omega t+\phi_{\zeta})}\Bigg\},\\
\end{equation}
while the displacement conductance $G^{d}_{\zeta}$ is given by
\begin{equation}
I^{d}(t)={\rm Re}\Bigg\{\displaystyle\sum\limits_{\zeta}G^{d}_{\zeta}(\Omega)v_{\zeta}^{\rm{ac}}e^{-i(\Omega t+\phi_{\zeta})}\Bigg\}.
\end{equation}

The conservation of the total charge current and gauge invariance with respect to the shift 
of the chemical potentials lead to $\sum_{\xi}G_{\xi\zeta}=0$ and $\sum_{\zeta}G_{\xi\zeta}=0$.\cite{But02}
These equations are satisfied by partitioning the displacement current into each lead,\cite{Guo1} $I^{\Omega, tot}_{\xi 0}=I^{\Omega}_{\xi 0}+A_{\xi}I^{d}$, 
or, equivalently, $G_{\xi\zeta}=G^{c}_{\xi\zeta}+A_{\xi}G^{d}_{\zeta}$, in such a way 
that the sum of the partitioning factors $A_{\xi}$ obeys $\sum_{\xi}A_{\xi}=1$. Using the sum rules given above one obtains the expression 
for the dynamic conductance,\cite{But02,Guo1}
\begin{equation}
	\label{eq:cond_part}
	G_{\xi\zeta}=G^{c}_{\xi\zeta}-G^{d}_{\zeta}\frac{\sum_{\lambda}G^{c}_{\xi\lambda}}{\sum_{\lambda}G^{d}_{\lambda}},
\end{equation} 
where $A_{\xi}=-(\sum_{\lambda}G^{c}_{\xi\lambda})/(\sum_{\lambda}G^{d}_{\lambda})$, $G^{d}_{\zeta}=-\sum_{\xi}G^{c}_{\xi\zeta}$, and
 $G(\Omega)=G_{LL}(\Omega)=G_{RR}(\Omega)=-G_{LR}(\Omega)=-G_{RL}(\Omega)$.
The first term of Eq.~(\ref{eq:cond_part}) represents the dynamic response of the charge current, 
while the second term is the internal response
to the applied external ac perturbation due to screening by Coulomb interaction.
Note that the dynamic conductance consists of a real dissipative component $G_{R}$, and 
an imaginary nondissipative component $G_{I}$ indicating the difference in phase between the current
and the voltage.
Due to the total current conservation, the two terms in Eq.~(\ref{eq:cond_part}) should behave in a way that a minimum (maximum) of $G^{c}_{\xi\zeta}(\Omega)$
corresponds to a maximum (minimum) of $G^{d}_{\zeta}(\Omega)$
for both real and imaginary parts.

\subsection{Density of states in the quantum dot}

Since the dynamic conductance is an experimentally directly accessible quantity, we hope that a measurement can help to reveal the internal time scales of the coupling between the molecular and electronic spins in the transport.
We begin by analyzing the 
density of states available for electron transport in the quantum dot
\begin{equation}
\rho(\epsilon)=-\frac{1}{\pi}\sum_{\sigma=\pm 1}{\rm Im}\bigg\{\frac{G^{0r}_{\sigma\sigma}(\epsilon)}{1-\gamma^{2}G^{0r}_{\sigma\sigma}(\epsilon)G^{0r}_{-\sigma-\sigma}(\epsilon_{\sigma})}\bigg\}.
\end{equation} 
There are four resonant transmission channels. They 
are positioned at quasienergy levels 
$\epsilon_{1}=\epsilon_{\downarrow}=\epsilon_{0}-(\omega_L + JS)/2$ (spin down), 
$\epsilon_{2}=\epsilon_{\downarrow}+\omega_L=\epsilon_{0}+(\omega_L - JS)/2$ (spin up), 
$\epsilon_{3}=\epsilon_{\uparrow}-\omega_L=\epsilon_{0}-(\omega_L - JS)/2$ (spin down) and
$\epsilon_{4}=\epsilon_{\uparrow}=\epsilon_{0}+(\omega_L + JS)/2$ (spin up).

The Hamiltonian of the molecular orbital is a periodic function of time $\hat{H}_{\textrm{MO}}(t)=\hat{H}_{\textrm{MO}}(t+\tau)$, with period $\tau=2\pi/\omega_{L}$. Its Fourier expansion is given by 
$\hat{H}_{\textrm{MO}}(t)=\sum_{n}\hat{H}^{(n)}_{\textrm{MO}}e^{in\omega_{L}t}$. Applying the Floquet theorem one can obtain the Floquet quasienergy $\epsilon_{\alpha}$ corresponding to the Floquet state $\lvert\psi_{\alpha}(t)\rangle$ in the Schr\"{o}dinger equation
\begin{equation}
\hat{\mathcal{H}}_{\textrm{MO}}(t)\lvert\psi_{\alpha}(t)\rangle=\epsilon_{\alpha}\lvert\psi_{\alpha}(t)\rangle,
\end{equation}
where $\hat{\mathcal{H}}_{\textrm{MO}}(t)=\hat{H}_{\textrm{MO}}(t)-i\partial_{t}$.\cite{Floquet1,Floquet2,Floquet3,Floquet4} The Floquet Hamiltonian matrix is block diagonal, with matrix elements given by $\langle\alpha;n\lvert\hat{H}_{F}\lvert\beta;m\rangle=[\hat{H}^{(n-m)}_{\textrm{MO}}]_{\alpha\beta}+n\omega_{L}\delta_{\alpha\beta}\delta_{nm}$,\cite{Floquet2} where $\lvert\alpha;n\rangle$ describes the Floquet states, while $\alpha$ denotes the electron spin states. For restricted Floquet quasienergies to the frequency interval $[0,\omega_{L})$ a block is given by 
\begin{equation}
\left (
\begin{array}{cc}
\lambda_{1}-\omega_{L} &  JS_{\bot}/2\\
\vspace*{-0.1cm}\\
JS_{\bot}/2 & \lambda_{2}
\end{array}
\right ),\label{eq: block}
\end{equation}
with $\lambda_{1,2}=\epsilon_{0}\pm (\omega_{L}+JS_{z})/2$. The corresponding Floquet quasienergies are eigenenergies of the matrix (\ref{eq: block}), equal to $\epsilon_{1}$ and $\epsilon_{3}$. The precessing component of the molecular spin couples states with quasienergies $\epsilon_{1}$ and $\epsilon_{3}$ to states with quasienerges $\epsilon_{2}$ and $\epsilon_{4}$, which differ in energy by an energy quantum $\omega_{L}$.
Namely, due to the periodic motion of the molecular spin an electron can absorb or emit an energy $\omega_L$, accompanied with a spin flip. Spin-flip processes due to rotating magnetic field were analyzed in some works.\cite{Guo,we} A similar mechanism was discussed in a recent work for a nanomechanical spin-valve, in which inelastic spin-flip processes are assisted by molecular vibrations.\cite{Pascal}

\subsection{Analysis of dynamic conductance}
\begin{figure*}[t]
	\includegraphics[height=5.7cm,keepaspectratio=true]{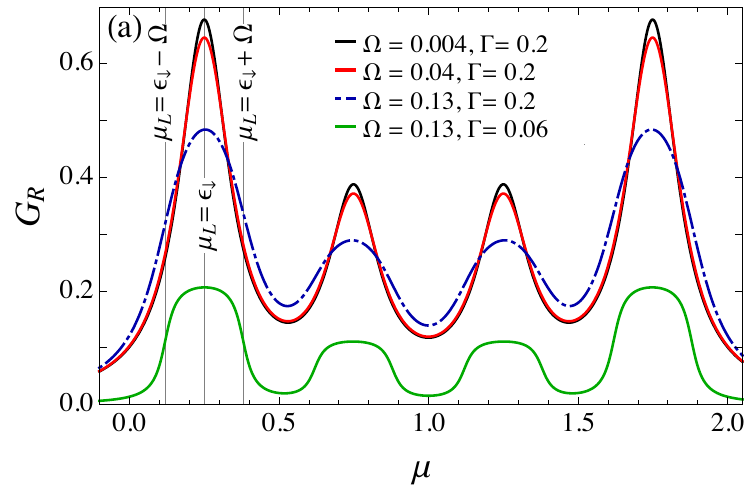}\,\,\,\,  
	\includegraphics[height=5.7cm,keepaspectratio=true]{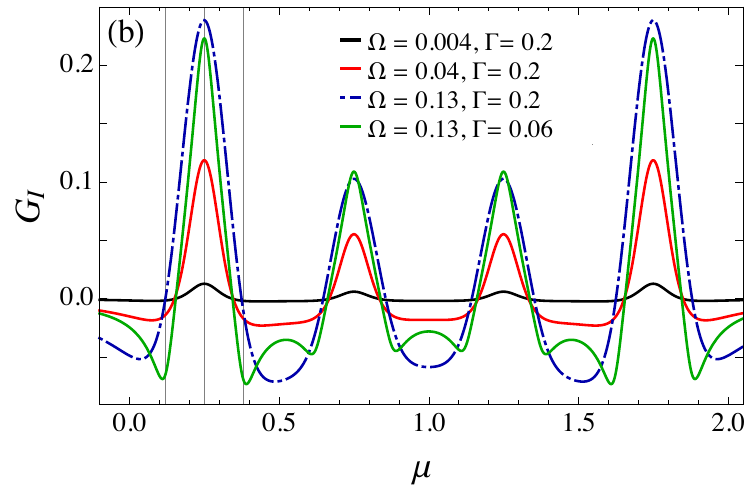}    
	\caption{(Color online) (a) Real part $G_{R}$ and (b) imaginary part
		$G_{I}$ of the dynamic conductance as functions of the chemical potential $\mu$, with $\mu=\mu_{L}=\mu_{R}$. 
		The plots are obtained for different ac frequencies $\Omega$ and tunneling rates $\Gamma$ at zero temperature, with $\Gamma_{L}=\Gamma_{R}=\Gamma/2$, and $\vec B=B\vec e_{z}$. All energies are given in the units of $\epsilon_{0}$. 
		The other parameters are set to: $\omega_L=\nobreak0.5,\, J=\nobreak0.01,\, S=100,\, \theta=1.25,\, \gamma\approx 0.474$. The molecular quasienergy levels are positioned at: $\epsilon_{1}=0.25$, $\epsilon_{2}=0.75$, $\epsilon_{3}=1.25$, and $\epsilon_{4}=1.75$.\,The conductance components $G_{R}$ and $G_{I}$ are given in the units of conductance quantum $e^{2}/h$.}\label{fig: conductance}
\end{figure*}
 
Now we analyze the charge conductance in response to the ac voltages. The suppression of dc conductance of charge current due to photon-assisted processes in the presence of an ac gate voltage, or a rotating magnetic field, was discussed in Ref.\citenum{Floquet4}. Here we consider ac conductance in a double-driving experiment, where we first induce molecular spin precession at Larmor frequency $\omega_{L}$, and then turn on the oscillating fields with frequency $\Omega$ in the leads. Assuming equal chemical potentials of the leads $\mu_{L}=\mu_{R}=\mu$, we analyze the dynamic conductance $G(\Omega)$ at zero temperature. Since we work in the wide-band limit,
this symmetry simplifies the partitioning factors to $A_{\xi}=\Gamma_{\xi}/\Gamma$. Hence, Eq.~(\ref{eq:cond_part}) can be transformed into
\begin{equation}
		\label{eq:final_conductance}
G_{\xi\zeta}(\Omega)=\frac{e^2}{h}\int d\epsilon T_{\xi\zeta}(\epsilon,\Omega) \frac{f_{\zeta}(\epsilon-\Omega)-f_{\zeta}(\epsilon)}{\Omega}.
\end{equation}
Here $T_{\xi\zeta}(\epsilon,\Omega)$ is the effective transmission function that can be expressed as 
$T(\epsilon,\Omega)=T_{LL}(\epsilon,\Omega)=T_{RR}(\epsilon,\Omega)=-T_{LR}(\epsilon,\Omega)=-T_{RL}(\epsilon,\Omega)$, which reads
\begin{widetext}
	\begin{equation}
		 \quad T(\epsilon,\Omega)=\frac{\Gamma_{L}\Gamma_{R}}{\Gamma}(\Gamma-i\Omega)\displaystyle\sum_{\sigma=\pm 1}\frac{G^{0r}_{\sigma\sigma}(\epsilon)G^{0a}_{\sigma\sigma}(\epsilon-\Omega)
			[1+\gamma^{2}G^{0r}_{-{\sigma}-{\sigma}}(\epsilon_{\sigma})G^{0a}_{-{\sigma}-{\sigma}}(\epsilon_{\sigma}-\Omega)]}
		{[1-\gamma^{2}G^{0a}_{\sigma\sigma}(\epsilon-\Omega)G^{0a}_{-{\sigma}-{\sigma}}(\epsilon_{\sigma}-\Omega)]
			[1-\gamma^{2}G^{0r}_{\sigma\sigma}(\epsilon)G^{0r}_{-{\sigma}-{\sigma}}(\epsilon_{\sigma})]}.
	\end{equation}
\end{widetext}

The real part $G_{R}$ and imaginary part $G_{I}$ of the dynamic conductance versus chemical potential $\mu$
are plotted in Figs.~\ref{fig: conductance}(a) and \ref{fig: conductance}(b). Both $G_R$ and $G_I$ achieve their maximum at $\mu_{\zeta}=\epsilon_{i}$, where
the resonance peaks are positioned. In accordance with Eq.~(\ref{eq:final_conductance})
the electrons in lead $\zeta=L,R$, with energies $\mu_{\zeta}-\Omega\leq\epsilon\leq\mu_{\zeta}$, can participate in the transport 
processes by absorbing a photon of energy $\Omega$. 
For $\Omega\rightarrow 0$ the dynamic conductance reduces to dc conductance $G_{\xi\zeta}(\Omega\rightarrow 0)=e^{2}T_{\xi\zeta}(\mu_{\zeta}, \Omega\rightarrow 0)/h$, and reaches its maximum at resonances given by the Floquet quasienergies.\cite{Floquet4}
The imaginary part of the dynamic conductance $G_I$ approaches zero for $\Omega\rightarrow 0$ [black line in Fig. \ref{fig: conductance}(b)]. 
The considerable contribution of the displacement current to the total current
is reflected in the decrease of $G_{R}$, and the increase of $G_{I}$ near resonances with increasing $\Omega$,
as the displacement current opposes the change of the particle charge current under ac bias [red and blue dot-dashed lines in Figs. \ref{fig: conductance}(a) and \ref{fig: conductance}(b)].
For a small value of both $\Gamma$ and $\Omega$, 
$G_{R}$ shows sharp resonant peaks. However, with the increase of $\Omega$, each of the peaks in $G_{R}$ 
broadens [green line in Fig. \ref{fig: conductance}(a)]. It approaches a constant value around the corresponding resonant level, with the width equal to $2\Omega$, since the inequality  
\begin{equation}
\label{eq:inequal}
\lvert\epsilon_{i}-\mu_{\zeta}\rvert\leq\Omega
\end{equation}
is the condition for the inelastic photon-assisted tunneling to occur. 
 	
\subsection{Frequency dependence of the ac conductance and equivalent circuit}\label{SecD}

\begin{figure} [h!]
	\includegraphics[height=4.0cm,keepaspectratio=true]{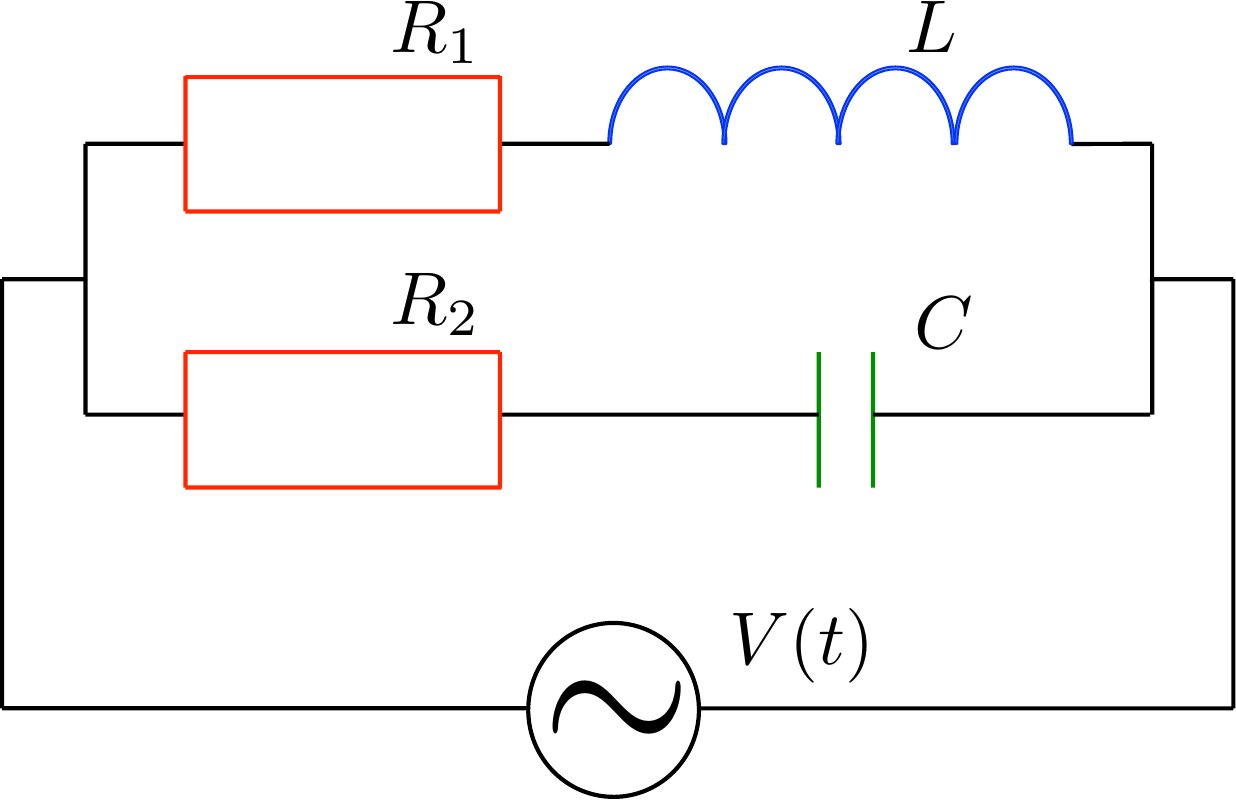}   
	\caption{(Color online) The equivalent classical circuit of the molecular magnet junction in the low-ac-frequency regime. It is composed of two serial combinations: one of a resistor and an inductor and the other of a resistor and a capacitor connected in parallel and driven by a source of ac voltage $V(t)$. The resistances are denoted by $R_1$ and $R_2$; $L$ is the inductance and $C$ is the capacitance of the circuit elements.}\label{fig: rlc}
\end{figure} 
\begin{figure*}
	\includegraphics[height=5.6cm,keepaspectratio=true]{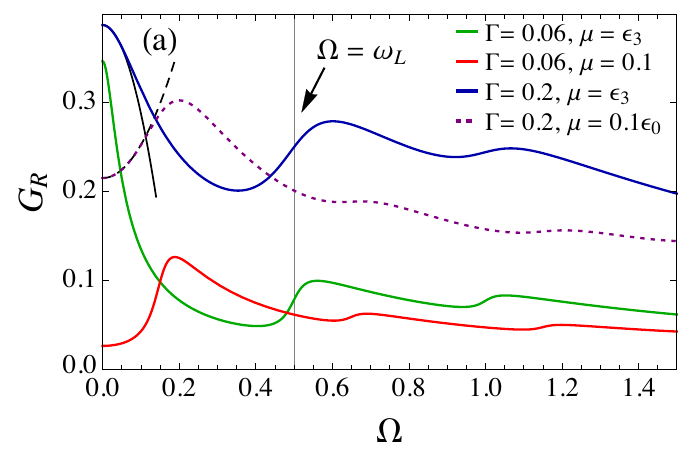}\,\,\,\includegraphics[height=5.6cm,keepaspectratio=true]{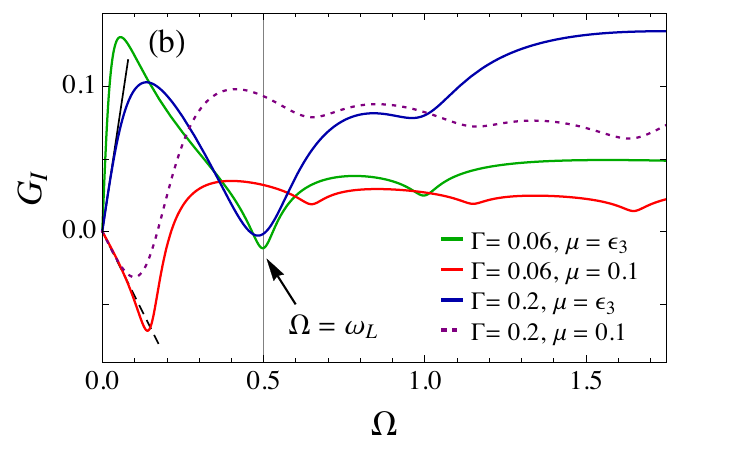}   
	\caption{(Color online) (a) Real part $G_R$ and (b) imaginary part $G_I$ of the dynamic conductance 
		as functions of the ac frequency $\Omega$.
		The plots are obtained for two different tunneling rates $\Gamma$ and chemical potentials $\mu$, with $\mu=\mu_{L}=\mu_{R}$ and $\vec{B}=B\vec{e}_{z}$, at zero temperature. All energies are given in the units of $\epsilon_{0}$. The other 
		parameters are set to: $\Gamma_{L}=\Gamma_{R}=\Gamma/2$, $S=100$, $J=0.01$, $\omega_{L}=0.5$, $\theta=1.25$, $\gamma\approx 0.474$. The molecular quasienergy levels lie at: $\epsilon_{1}=0.25$, $\epsilon_{2}=0.75$, $\epsilon_{3}=1.25$, and $\epsilon_{4}=1.75$. In the resonant case $\mu=\epsilon_3$, the response of the system is inductive-like in the low-ac-frequency limit ($G_{I}>0$), and $G_R$ and $G_I$ are both enhanced around $\Omega=\omega_L$, after going to a local minimum, as the channel with quasienergy $\epsilon_4$ becomes available for photon-assisted tunneling, i.e., $\mu+\Omega=\epsilon_4$. The conductance components $G_{R}$ and $G_{I}$ are given in the units of $e^{2}/h$.
		}\label{fig: frequency}
\end{figure*}
The behavior of the ac-conductance in the low-ac-frequency regime can be understood  using a classical circuit theory.\cite{But004} 
Namely, at small ac frequencies $\Omega\ll\Gamma$, the molecular magnet junction behaves 
as a parallel combination of two serial connections: one of a resistor and an inductor and the other
of a resistor and a capacitor, i.e., as a classical electric circuit (see Fig.~\ref{fig: rlc}).
Depending on the phase difference between the voltage and the current, the circuit shows
inductive-like (positive phase difference) or capacitive-like (negative phase difference) responses to the applied ac voltage. 
Thus, the dynamic conductance can be expanded
up to the second order in $\Omega$ in the small-ac-frequency limit as
\begin{align}
	\label{eq:approx_conductance}
G(\Omega)&= G(0)+G'(0)\Omega+\frac{1}{2}G''(0)\Omega^2 +O(\Omega^3)\nonumber\\
&\approx\frac{1}{R_1}+i\left(\frac{L}{R_1^2}-C\right)\Omega+\left(R_{2}C^2-\frac{L^2}{R_1^3}\right)\Omega^{2},
\end{align}
where $R_1$, $R_2$, L, and C denote the resistances, inductance and capacitance of the circuit.
In our further analysis we will assume that $R_{1}=R_{2}=R$.
The first term of Eq.~(\ref{eq:approx_conductance}) represents the dc conductance $G(0)=1/R$. The second, imaginary term,
linear in $\Omega$, is $iG_{I}$ in the low-ac-frequency limit. 

Depending on the sign of
$L/R^2-C$, the linear response is inductive-like ($G_{I}>0$) while $G_R$ decreases, or 
capacitive-like ($G_{I}<0$) while $G_R$ increases with the increase of $\Omega$. 
For $C=L/R^2$ the system behaves like a resistor with $G=G(0)$.
The nondissipative component $G_I$ shows inductive-like behavior for 
\begin{equation}
\label{eq:inequal2}	
\rvert\epsilon_{i}-\mu_{\zeta}\rvert<\frac{\Gamma}{2},
\end{equation}
as we have observed in Fig.~\ref{fig: conductance}(b) (red line), and capacitive-like or resistive behavior otherwise.

The behavior of the dynamic conductance components $G_{R}$ and $G_{I}$ as functions of the ac frequency $\Omega$ for $\mu=\epsilon_3$ 
and $\mu=0.1\,\epsilon_{0}$, with two values of $\Gamma$ at zero temperature is presented in Fig.~\ref{fig: frequency}. The real part $G_R$ is an even,
while the imaginary part $G_I$ is an odd function of $\Omega$.
In the low-ac-frequency regime $\Omega\ll\Gamma$, $G_{R}$ is a quadratic function, while $G_{I}$ is a linear function of ac frequency (solid and dashed black lines in Fig.~\ref{fig: frequency}). 
By fitting parameters of these functions and using Eq.~(\ref{eq:approx_conductance}), one obtains circuit parameters $R$, $L$, and $C$, confirming
that in this limit the ac conductance of the system resembles the previously described classical circuit model.
The circuit parameters can be calculated in terms of the dynamic conductance according to Eq.~(\ref{eq:approx_conductance}).
Note that they depend on the relative position of the Fermi energy of the leads with respect to the molecular quasienergy levels.

Near the four resonances we expect the system to be highly transmissive and therefore to conduct well. 
This is confirmed by Figs.~\ref{fig: conductance} and \ref{fig: frequency}. Namely, the imaginary conductance component $G_{I}>0$
around resonances and is a positive 
linear function of $\Omega$ in the low-ac-frequency limit [see Fig.~\ref{fig: frequency}(b), black solid line]. This implies that
the behavior of the system is inductive-like,
since the displacement current tends to reduce the charge current,
as electrons reside awhile in the quantum dot, causing the delay in phase between the voltage and the current.
Accordingly, the real component $G_{R}$ decreases quadratically from initial value $G(0)$ upon switching on the ac frequency $\Omega$ [black solid line in Fig.~\ref{fig: frequency}(a)].
However, the off-resonance behavior is capacitive-like resulting from 
intraorbital Coulomb interactions, included via displacement current.\cite{Guo1}
Hence, in the low-ac-frequency limit $G_{I}(\Omega)$ is negative and decreases linearly with the increase of $\Omega$ for Fermi energies 
of the leads which are far from the resonant energies $\epsilon_{i}$ [black dashed line in Fig.~\ref{fig: frequency}(b)]. In this case $G_{R}(\Omega)$ increases quadratically with $\Omega$ [black dashed line in Fig.~\ref{fig: frequency}(a)]. Obviously, the molecular magnet junction behaves as a classical circuit only in the low-ac-frequency regime.

For higher ac frequencies $\Omega$ we 
use Eq.~(\ref{eq:final_conductance}) to analyze the behavior of $G_R$ and $G_{I}$, where
the dynamic response of the system remains predominantly inductive-like for $\mu=\epsilon_{\uparrow}-\omega_{L}=\epsilon_3$.
With further increase of $\Omega$, the ac conductance $G(\Omega)$ vanishes asymptotically. Upon turning on the ac frequency, while 
the system is on resonance $\mu=\epsilon_{\uparrow}-\omega_L$, the imaginary 
component $G_{I}$ increases quickly from 0 to a local maximum
and then decreases to its minimum value around $\Omega=\omega_{L}$ [green and blue lines in Fig.~\ref{fig: frequency}(b)]. The real part $G_{R}$ decreases to a local minimum 
and then has a steplike increase towards a local maximum around $\Omega=\omega_{L}$ [green and blue lines in Fig. \ref{fig: frequency}(a)]. 
This behavior of the dynamic conductance can be understood as follows.
For $\mu=\epsilon_{\uparrow}-\omega_L$, at $\Omega=\omega_{L}$, besides the resonant level with quasienergy $\epsilon_{\uparrow}-\omega_{L}$, the upper level
with quasienergy $\epsilon_{\uparrow}$ becomes available for photon-assisted electron transport. It is then distanced by the energy $\Omega$ from the chemical potential $\mu$. Consequently, an electron
with Fermi energy equal to $\epsilon_{\uparrow}-\omega_{L}$ can absorb a photon of energy $\Omega=\omega_{L}$ in the lead $\zeta$ and tunnel into the level with 
quasienergy $\epsilon_{\uparrow}$. 
This leads to an enhancement of the response functions $G_{R}$ and $G_{I}$, after going to a local minimum, with features corresponding to photon-assisted tunneling processes. Each steplike increase of $G_R$ and the corresponding dip of $G_I$
in Fig.~\ref{fig: frequency} are determined by the difference between the quasienergy levels $\epsilon_{i}$ and the chemical potential $\mu$, viz.~$\rvert\epsilon_{i}-\mu\rvert=\Omega$.
Thus, for $\mu=\epsilon_3$ and the set of parameters given in Fig.~\ref{fig: frequency}, they are positioned around $\Omega/\epsilon_{0}=0.5$ 
and $\Omega/\epsilon_{0}=1$. For the larger tunnel couplings each steplike increase in $G_R$ is broadened due to the level broadening $\Gamma$. We notice that the enhancement of the dynamic conductance is higher around $\Omega=\omega_L$ than around the subsequent frequency $\Omega/\epsilon_{0}=1$. This is due to the fact that the frequency has to
traverse one resonant peak in $G_R$, or dip in $G_I$, to reach the second one.
We need to mention that the off-diagonal conductances $G_{\xi\zeta}=-G$, where $\xi\neq\zeta$, and hence
have a behavior that opposes that of the diagonal ones.

\begin{figure*}
	\includegraphics[height=5.4cm,keepaspectratio=true]{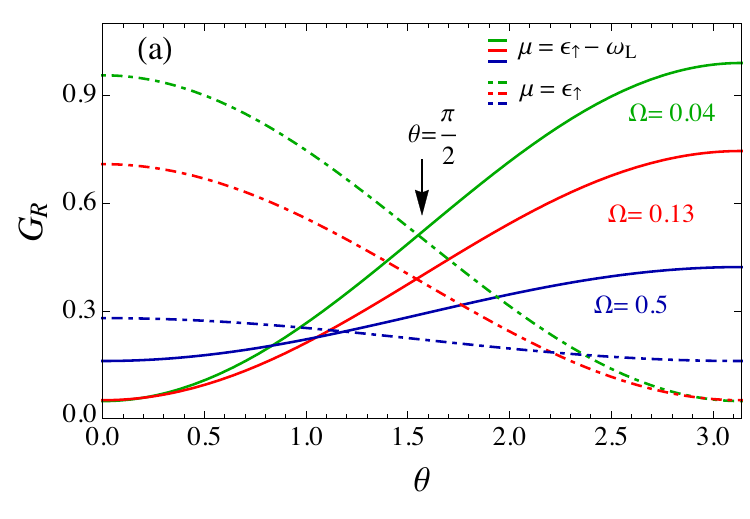}\,\,\,\,
	\includegraphics[height=5.4cm,keepaspectratio=true]{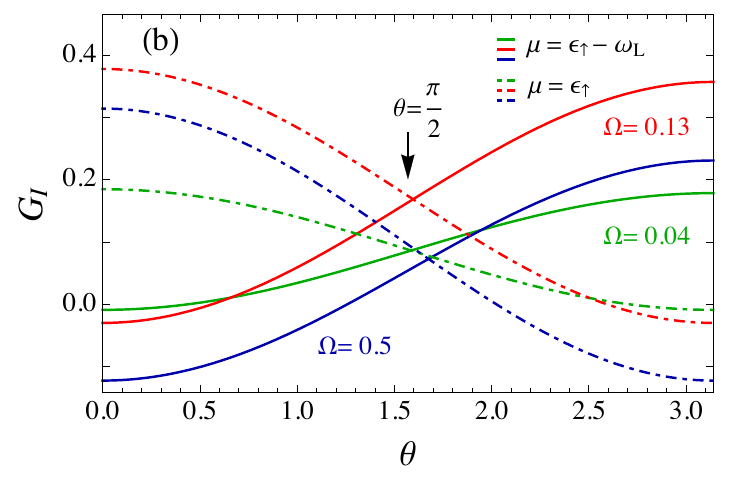}   
	\caption{(Color online) (a) Real part $G_R$ and (b) imaginary part $G_I$ of the dynamic conductance 
		as functions of the tilt angle $\theta$ of the molecular spin $\vec{S}$ from the magnetic field $\vec{B}=B \vec e_{z}$.
		The plots are obtained for different values of 
		$\Omega$ and $\mu$, with $\mu=\mu_{L}=\mu_{R}$, at zero temperature. All energies are given in the units of $\epsilon_{0}$. The other 
		parameters are set to $S=100$, $J=0.01$, $\omega_{L}=0.5$, $\Gamma=0.2$,
		and $\Gamma_{L}=\Gamma_{R}=\Gamma/2$. In the limit of low frequency $\Omega$, for $\theta\rightarrow\pi/2$, the conductance component $G_{R}$, as well as $G_I$, approaches equal value at each resonance. The conductance components $G_{R}$ and $G_{I}$ are given in the units of conductance quantum $e^{2}/h$.}\label{fig: theta}
\end{figure*}
In the spirit of the scattering matrix formalism, the dynamic conductance of our molecular magnet junction, 
in the low-ac-frequency regime, can be expanded as\cite{But04} 
\begin{equation}
\label{eq:emittance}
G_{\xi\zeta}(\Omega)=G_{\xi\zeta}(0)-i\Omega E_{\xi\zeta}+\Omega^{2}K_{\xi\zeta}+O(\Omega^3), 
\end{equation}
where
$G_{\xi\zeta}(0)$ is the dc conductance. The quantity $E_{\xi\zeta}=-{\rm Im}\{ \partial G_{\xi\zeta}(0)/\partial\Omega\}$ is called the emittance.\cite{But04} It contains the contribution
from the displacement current and the partial density of states that characterize 
the scattering process.\cite{But03,But05,But06} The partial density of states can be calculated using the scattering matrix,
and can be understood as density of states due to electrons injected from lead $\zeta$, and leaving through lead $\xi$.\cite{But03,But05,But06}
The emittance $E_{\xi\zeta}$ measures the dynamic response of the system to an external oscillating ac field
and, depending on its sign, the response is capacitive-like or inductive-like.\cite{But04}
The matrix element of the third term, $K_{\xi\zeta}={\rm Re}\{ \partial^{2}G_{\xi\zeta}(0)/\partial\Omega^{2}\}/2$, represents the correction to the real 
part of the dynamic conductance and describes the dynamic dissipation
in the low-ac-frequency regime.\cite{But04} Both $E_{\xi\zeta}$ and $K_{\xi\zeta}$ obey the sum rules, since the
total current conservation and gauge invariance conditions have to be satisfied.\cite{But02} According to Eq.~(\ref{eq:emittance}), their diagonal 
elements $E=E_{\xi\xi}$ and $K=K_{\xi\xi}$ can be approximated as $E\approx -G_{I}/\Omega$ and 
$K\approx[G_{R}-G(0)]/\Omega^2$ in the low frequency limit.\cite{But04} Based on the analyzed $G_{R}$ and $G_{I}$
the behavior of $E$ and $K$ can be examined.
Around all resonances $\mu=\epsilon_{i}$ the emittance $E<0$ (inductive-like response) and $K<0$ since $G_{R}<G(0)$, 
while off resonance $E>0$ (capacitive-like response) and $K>0$ (see Figs.~\ref{fig: conductance} and \ref{fig: frequency}).
 
\subsection{Effects of the molecular magnetization direction on the ac conductance}

Now we analyze the ac conductance components $G_{R}$ and $G_{I}$ as functions of the tile angle $\theta$ of the molecular spin $\vec{S}$ from the external field $\vec{B}$, plotted in Figs.~\ref{fig: theta}(a) and \ref{fig: theta}(b). For $\theta=1.25$, the peaks of both $G_{R}$ and $G_{I}$ in Figs.~\ref{fig: conductance}(a) and \ref{fig: conductance}(b) at $\mu=\epsilon_{\uparrow,\downarrow}\mp\omega_{L}$ 
are much lower than those at $\mu=\epsilon_{\uparrow,\downarrow}$, implying that the molecular magnet junction is less transmissive
at the upper two mentioned resonances. This can be qualitatively understood by looking at Fig.~\ref{fig: theta}. 
The behavior of the conductance components near the resonances for $\mu=\epsilon_{\uparrow}-\omega_L$ (solid lines in Fig.~\ref{fig: theta})
and $\mu=\epsilon_{\uparrow}$ (dot-dashed lines in Fig.~\ref{fig: theta}),
depends on the direction of $\vec{S}$ with respect to the external magnetic field $\vec{B}$. 
For $\theta=0$ the molecular spin $\vec{S}$ is static and the only two levels available for electron 
transport are Zeeman levels $\epsilon_{1}=\epsilon_{\downarrow}$ and $\epsilon_{4}=\epsilon_{\uparrow}$.
In this case, when the system is at the resonance $\mu=\epsilon_{\uparrow}$,  
the components $G_{R}$ and $G_{I}$ take their maximum values, and $G_{I}>0$ displaying an inductive-like behavior.
For $\mu=\epsilon_{\uparrow}-\omega_L$ and $\theta=0$,  
both $G_{R}$ and $G_{I}$ take their minimum values. There is no transmission 
channel at this energy for $\theta=0$, but $\Gamma$ is relatively large,
and $G_{I}<0$ displays a capacitive-like response. With the increase of $\theta$, the additional two channels
at energies $\epsilon_{\uparrow}-\omega_{L}$ and $\epsilon_{\downarrow}+\omega_L$ appear and become available for electron transport. This leads to
the increase of conductance components $G_{R}$ and $G_{I}$ at $\mu=\epsilon_{\uparrow}-\omega_L$, and their decrease  at $\mu_{L}=\epsilon_{\uparrow}$, as functions of $\theta$ (see Fig.~\ref{fig: theta}).
For $\theta\rightarrow\pi/2$, in the case of small $\Omega$ the complex components of the effective transmission function $T(\epsilon,\Omega)$
approach the same height at resonant energies $\epsilon_{i}$, so the probability of transmission reaches
equal value at each level. Thus, both $G_{R}$ and $G_{I}$ show peaks of the same height at the resonances. The points of intersection of solid and 
dot-dashed lines of the same color in Fig.~\ref{fig: theta} correspond to this particular case. For larger frequencies $\Omega$, these points are shifted away from $\theta\rightarrow\pi/2$, since the peaks broaden and overlap and the suppression or increase of $G_R$ and $G_I$ is much faster.
Finally, for $\theta=\pi$ the situation is reversed compared to the one with $\theta=0$, as
again the static spin $\vec{S}$ is in the direction opposite that of the external field $\vec{B}$. The Zeeman splitting in this case is equal to $\omega_{L}-JS$,
so the only two levels available for electron transport are $\epsilon_{2}$ and $\epsilon_{3}$.
Therefore, for $\theta=\pi$, when the system is at the resonance $\mu=\epsilon_{3}$, the conductance components $G_{R}$ and $G_{I}$ reach their maximum values, with $G_{I}>0$.
For $\mu=\epsilon_{4}$, which is off resonance for $\theta=\pi$, both $G_{R}$ and $G_{I}$ take minimum values, with $G_{I}<0$.

\section{Spin transport and spin-transfer torque}
\subsection{Spin transport under dc-bias voltage}

In the absence of ac harmonic potentials in the leads, tunneling under dc-bias voltage takes place.
The spin-angular momenta between the itinerant electronic spins and the precessing molecular spin are exchanged via 
exchange interaction, governed by the coupling constant $J$.
The molecular spin precession pumps spin currents into the system, but remains undamped using external sources, which compensate effects of the interaction with electron spins.
Further simplification of Eq.~(\ref{eq:prec_ind_current}) 
gives time-independent $z$ component of the spin current, $I_{Lz}^{\omega_{L}}$, and the in-plane $j=x,y$ 
time-dependent spin-current components from the left lead
\begin{equation}
 I^{\omega_L}_{L j }(t)=[I_{Lj}(\omega_{L}){e^{-i\omega_{L}t}}+I^{*}_{Lj}(\omega_L){e^{i\omega_{L}t}}].\label{eq: k resonant}
\end{equation}
 The expressions for complex time-independent functions $I_{Lx}(\omega_L)$ and $I_{Ly}(\omega_L)$, and the spin current $I_{Lz}^{\omega_{L}}$ are given by Eqs.~(\ref{eq: x resonant})--(\ref{eq: z resonant}) in the Appendix.

The spin-transport properties are characterized by elastic, i.e., energy-conserving tunnel processes [terms involving factors $[f_{L}(\epsilon)-f_{R}(\epsilon)]$ in Eqs.~(\ref{eq: x resonant}) and (\ref{eq: z resonant})], and inelastic, i.e., energy-nonconserving tunnel processes [terms involving factors $[f_{\xi}(\epsilon-\omega_{L})-f_{\zeta}(\epsilon)]$ in Eqs.~(\ref{eq: x resonant}) and (\ref{eq: z resonant})]. In the later ones an electron changes its energy 
by $\omega_L$ and flips its spin due to the exchange interaction with the rotational component of the molecular spin.
The spin-flip processes occur between levels with quasienergies $\epsilon_\uparrow$ and $\epsilon_{\uparrow}-\omega_L$ and between levels with quasienergies $\epsilon_\downarrow$ and
$\epsilon_{\downarrow}+\omega_L$.

The STT exerted by the inelastic spin-currents onto the spin of the molecule is given by\cite{slonc,berg,tser,ralp}
\begin{eqnarray}\label{eq: T resonant}
\vec {T}^{\omega_L}(t)=&-[{\vec I}^{\omega_L}_L(t)+{\vec I}^{\omega_L}_R(t)],
\end{eqnarray}
and can be expressed in
terms of the matrix elements of the Green's functions $\hat{G}^{0r}(\epsilon)$ and $\hat{G}^{0a}(\epsilon)$ as
\begin{align}
T^{\omega_{L}}_{j}(t)=\label{eq: torque Tk}&-\int\frac{d\epsilon}{2\pi}\sum_{\xi\zeta}\frac{\Gamma_{\xi}
\Gamma_{\zeta}}{\Gamma}[f_{\xi}(\epsilon-\omega_L)-f_{\zeta}(\epsilon)]\nonumber\\
&\times{\rm Im}\bigg\{({{\hat\sigma}}_{j})_{21}\frac{\gamma G^{0r}_{11}(\epsilon)G^{0a}_{22}(\epsilon-\omega_L)}
{\lvert 1-\gamma^{2}G^{0r}_{11}(\epsilon)G^{0r}_{22}(\epsilon-\omega_{L})\lvert^{2}}\nonumber\\
&\times [1-\gamma^{2}G^{0a}_{11}(\epsilon)G^{0r}_{22}(\epsilon-\omega_L)]e^{-i\omega_{L} t}\bigg\},\\
T^{\omega_L}_{z}=\label{eq: torque Tz}&-\int\frac{d\epsilon}{2\pi}\sum_{\xi\zeta}\Gamma_{\xi}
\Gamma_{\zeta}[f_{\xi}(\epsilon-\omega_L)-f_{\zeta}(\epsilon)]\nonumber\\
&\times\frac{\gamma^{2}\lvert G^{0r}_{11}(\epsilon) G^{0r}_{22}(\epsilon-\omega_L)\lvert^2}
{\lvert 1-\gamma^{2}G^{0r}_{11}(\epsilon)G^{0r}_{22}(\epsilon-\omega_{L})\lvert^{2}}.
\end{align} 
Regarding the moecular spin $\vec{S}$, the STT can be presented as
\begin{eqnarray}
\vec T^{\omega_L}(t)=\frac{\alpha}{S}\dot{\vec{S}}(t)\times\vec{S}(t)+\beta\dot{\vec{S}}(t)+\eta\vec{S}(t),\label{eq: torque-Gilbert}
\end{eqnarray}
with the Gilbert damping coefficient $\alpha$ in the first term. The coefficient $\beta$ that characterizes the modulation
of the precession frequency of the molecular spin $\vec{S}(t)$ is given by the second term. The third coefficient $\eta$
can be written in terms of $\alpha$ and $T^{\omega_{L}}_{z}$ as $\eta=[{T^{\omega_{L}}_{z}+\omega_{L} S\alpha\sin^{2}(\theta)}]/{S_z}$.
Using Eqs.~(\ref{eq: torque Tk}) and (\ref{eq: torque Tz}), and comparing them with Eq.~(\ref{eq: torque-Gilbert}), one obtains exact expressions for the torque coefficients $\alpha$ and $\beta$ as
\begin{figure*}[t]
	\includegraphics[height=4.95cm,keepaspectratio=true]{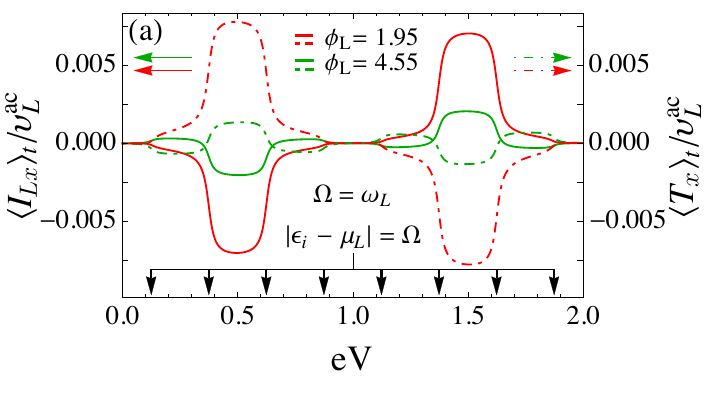}\,\,\,  
	\includegraphics[height=4.95cm,keepaspectratio=true]{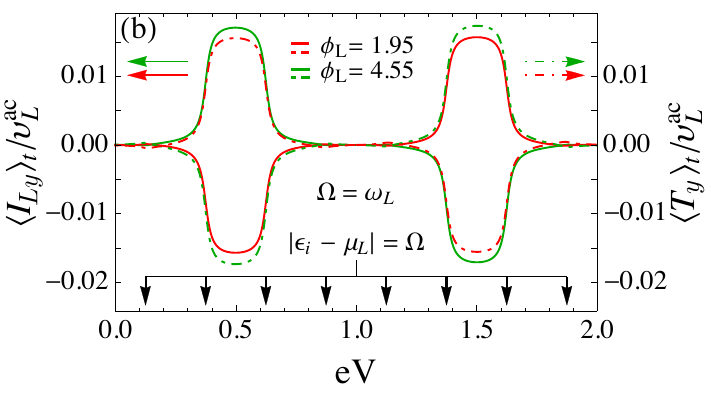}    
	\caption{(Color online) Bias-voltage dependence of the time-averaged components of the spin-current and spin-transfer torque (a) $\langle I_{Lx}\rangle_{t}/v^{\mathrm{ac}}_{L}$ and $\langle T_{x}\rangle_{t}/v^{\mathrm{ac}}_{L}$, and (b) $\langle I_{Ly}\rangle_{t}/v^{\mathrm{ac}}_{L}$ and $\langle T_{y}\rangle_{t}/v^{\mathrm{ac}}_{L}$.
		The plots are obtained at zero temperature for two different phases $\phi_L$, with $\vec B=B\vec e_{z}$. All energies are given in the units of $\epsilon_0$. The other 
		parameters are set to
		$\Gamma=0.04$, $\Gamma_{L}=\Gamma_{R}=\Gamma/2$,
		$ \mu_{R}=0,\,\phi_{R}=0$,
		$v^{\textrm{ac}}_{R}=0, \, 
		\theta=1.25, \,S=100, \,J=0.01$, and
		$\Omega=\omega_{L}=0.25$. Photon-assisted spin transport is enhanced for $\epsilon_{1}<\mu_{L}<\epsilon_{2}$ and $\epsilon_{3}<\mu_{L}<\epsilon_{4}$, where the in-plane components of the spin-current and spin-transfer torque approach the constant largest magnitudes.}\label{fig: in-plane average}
\end{figure*} 
\begin{widetext}
\begin{align}
\alpha=\label{eq: alpha}&-\frac{1}{\omega_{L}S}\int\frac{d\epsilon}{2\pi}\sum_{\xi\zeta}\Gamma_{\xi}
\Gamma_{\zeta}[f_{\xi}(\epsilon-\omega_L)-f_{\zeta}(\epsilon)]
\frac{( JS_{z}/2\Gamma){\rm Im}\{G^{0r}_{11}(\epsilon)G^{0a}_{22}(\epsilon-\omega_L)\}-\gamma^{2}\lvert G^{0r}_{11}
(\epsilon)G^{0r}_{22}(\epsilon-\omega_L)\lvert^2}
{\lvert 1-\gamma^{2}G^{0r}_{11}(\epsilon)G^{0r}_{22}(\epsilon-\omega_{L})\lvert^{2}},\\
\beta=\label{eq: beta}&-\frac{J}{\omega_{L}}\int\frac{d\epsilon}{4\pi}\sum_{\xi\zeta}\frac{\Gamma_{\xi}
\Gamma_{\zeta}}{\Gamma}[f_{\xi}(\epsilon-\omega_L)-f_{\zeta}(\epsilon)]
\frac{{\rm Re}\{G^{0r}_{11}(\epsilon)G^{0a}_{22}(\epsilon-\omega_L)\}-\gamma^{2}\lvert G^{0r}_{11}(\epsilon) G^{0r}_{22}(\epsilon-\omega_L)\lvert^2}
{\lvert 1-\gamma^{2}G^{0r}_{11}(\epsilon)G^{0r}_{22}(\epsilon-\omega_{L})\lvert^{2}}.
\end{align} 
\end{widetext}
In the limit $\gamma^{2}\rightarrow 0$, the expressions (\ref{eq: torque Tk})--(\ref{eq: beta}) are in agreement with Ref. \citenum{we}. In the strong exchange coupling limit $J\gg\Gamma$ both Gilbert damping coefficient $\alpha$ and the torque coefficient $\beta$ drop to zero.

\subsection{Photon-assisted spin transport under ac-bias voltage}\label{sec: PA}
We consider spin transport in the double-driving experiment, where we first establish molecular spin precession at Larmor frequency $\omega_{L}$ and then apply the oscillating potentials with frequency $\Omega$ in the leads. The spin current components indicating photon-assisted inelastic spin transport can be obtained by further simplification of Eq.~(\ref{eq:ac_voltage_current}).
The in-plane $x$ and $y$ spin-current components consist of oscillating terms involving both ac frequency $\Omega$ and Larmor frequency $\omega_L$. Experimentally, by adjusting $\Omega=\pm\omega_L$, these currents may be measurable. In this case they have one dc component and one component oscillating with frequency $2\Omega$. The photon-assisted spin currents are given by Eqs.~(\ref{eq: k assisted})--(\ref{eq: z assisted}) in the Appendix.

The time average of a periodic function $F(t)$, with a period T is defined as
\begin{eqnarray}
 \langle F\rangle_{t}=\frac{1}{T}\int^{T}_{0}F(t)dt.
\end{eqnarray}
According to Eq. (\ref{eq: k assisted}), the time-averaged $j=x,y$ components of the total spin current $\vec{I}_{L}(t)$ are nonzero only for $\Omega=\pm\omega_L$ and read
\begin{equation}
 \langle I_{Lj}\rangle_{t}=\langle I^{\Omega=\pm\omega_{L}}_{Lj}\rangle_{t}=
 \sum_{\xi}{\rm Re}\big\{I^{j}_{L\xi}(-\omega_{L})e^{\pm i\phi_{\xi}}\big\},\label{eq: x-y-aver} 
 \end{equation}
 while the time-averaged $z$ component of the spin-current equals
 \begin{equation}
 \langle I_{Lz}\rangle_{t}=
 I^{\omega_L}_{Lz}.
\end{equation}
Hence, the in-plane time-averaged $x$ and $y$ spin-current components contain only contributions from photon-assisted spin tunneling processes, while the $z$ component contains only contributions from spin tunneling under dc-bias voltage.
The time-averaged STT is then given by
\begin{equation}
\langle\vec{T}\rangle_{t}=-\sum_{\xi}\langle\vec{I_{\xi}}\rangle_{t}.
\end{equation} 
All the torques are compensated by external means, which keep the molecular spin precession undamped during the experiment. 

\subsection{Analysis of the time-averaged spin transport}

The in-plane components of the time-averaged spin current and STT are presented in Figs.~\ref{fig: in-plane average}(a) and \ref{fig: in-plane average}(b) as functions of the bias-voltage $eV=\mu_{L}-\mu_{R}$. According to Eqs.~(\ref{eq: y assisted}) and (\ref{eq: x-y-aver}), $\langle I_{Lx}\rangle_{t}$ and $\langle I_{Ly}\rangle_{t}$ differ in phase by $\pi/2$.
The plots are obtained at zero temperature for two different phases of ac field in the left lead. We set $\Omega=\omega_L$, the right lead's Fermi energy $\mu_{R}=0$, and apply an ac harmonic chemical potential only to the left lead. 
According to the segment $[f_{L}(\epsilon-\Omega)-f_{L}(\epsilon)]$ in Eq. (\ref{eq: x assisted}), electrons with energies within the window $[\mu_{L}-\Omega,\,\mu_{L}]$ participate in the photon-assisted spin transport. 
Each of these processes is followed by a spin-flip and emission (apsorbtion) of an amount of energy $\omega_{L}$. This is caused by the interaction of the electron spin with the precessing component of the molecular spin. 
In turn, during the exchange interaction, a photon-assisted STT is generated onto $\vec{S}(t)$.
In regard to photon-assisted transmission of $1/2$-spin particles, the in-plane spin-current components show significant changes in either magnitude or direction, controlled by the change of the phase of the ac field in the left lead $\phi_L$.
Similarly to the case of charge transport, the necessary condition for photon-assisted spin tunneling
is given by the inequality (\ref{eq:inequal}).
The cases with equality sign in (\ref{eq:inequal}) are  represented by the black arrows in Fig.~\ref{fig: in-plane average}, pointing to the $eV$ scale. Each level satisfying this condition corresponds to two black arrows. In the region between each two black arrows the inequality (\ref{eq:inequal}) is satisfied for at least one molecular quasienergy level. Here the components of spin current and STT approach constant values. 
If  $\epsilon_{1}\leq\mu_{L}\leq\epsilon_{2}$ or $\epsilon_{3}\leq\mu_{L}\leq\epsilon_{4}$, then
the inequality (\ref{eq:inequal}) is satisfied for both $\epsilon_{1}$ and $\epsilon_{2}$, or $\epsilon_{3}$ and $\epsilon_{4}$. As a result, the magnitude of spin currents and STT is enhanced under these conditions, due to the involvement of both levels $\epsilon_1$ and $\epsilon_2$, or $\epsilon_3$ and $\epsilon_4$, in photon-assisted spin transport and photon-assisted spin-flip processes. 
We should point out that both spin-current components and STTs are antisymmetric functions of $eV$ with respect to the position of $\epsilon_{0}$. This is a consequence of the antisymmetric position of levels $\epsilon_{i}$ attributed to spin-up or spin-down state of the electron with respect to $\epsilon_0$. Using Eq.~(\ref{eq: x-y-aver}) with $v_{R}^\mathrm{ac}=0$, $\phi_{R}=0$, we obtain the largest magnitudes of the $j=x,y$ time-averaged spin-current components for 
\begin{figure}
	\includegraphics[height=5.5cm,keepaspectratio=true]{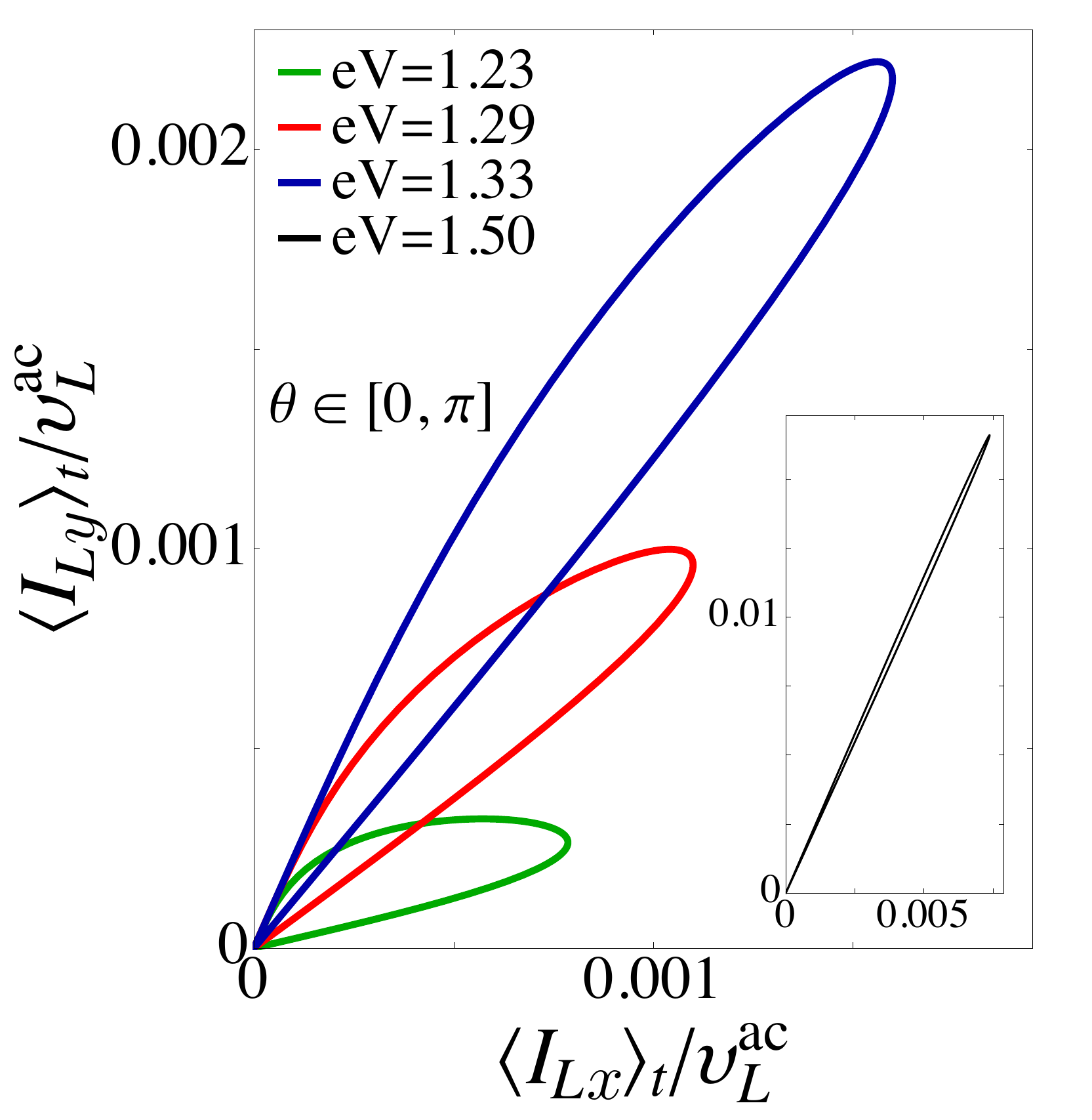}  
	\caption{(Color online) Time-averaged spin currents $\langle I_{Lx}\rangle_{t}/v^{\mathrm{ac}}_{L}$ and $\langle I_{Ly}\rangle_{t}/v^{\mathrm{ac}}_{L}$ for $\theta\in [0,\pi]$ and  $\phi_{L}=1.95$, with maximal magnitudes around $\theta=\pi/2$. The plots are obtained at zero temperature for four different bias-voltages varied as $eV=\nobreak\mu_{L}$. All energies are given in the units of $\epsilon_0$. The other parameters are the same as in Fig.~\ref{fig: in-plane average}. Inset: In-plane spin-current components for $eV=1.5$, where $\epsilon_{3}<\mu_{L}<\epsilon_{4}$.}\label{fig: parametric}
\end{figure}
\begin{equation}
\phi_{L}=\arctan\Bigg (\frac{{\rm Im}\{I^{j}_{LL}(-\omega_{L})\}}{{\rm Re}\{I^{j}_{LL}(-\omega_{L})\}}\Bigg ).
\end{equation} 
Simultaneously, the other in-plane time-averaged spin-current equals zero.
 
The magnitude of the time-averaged spin currents (and STTs) can also be controlled by tuning the tilt angle $\theta$, as presented in Fig.~\ref{fig: parametric}. For $\theta=0$, the in-plane spin currents are equal to 0. 
Since the spin-flip is most probable with the largest magnitude of the rotating field, the maximal magnitudes of $\langle I_{Lx}\rangle_{t}$ and $\langle I_{Ly}\rangle_{t}$
are obtained around $\theta=\pi/2$ and increase significantly if $\mu_L$ lies between any two levels connected with spin-flip mechanism (see the inset in Fig.~\ref{fig: parametric}). 

\begin{figure}
	\includegraphics[height=2.6cm,keepaspectratio=true]{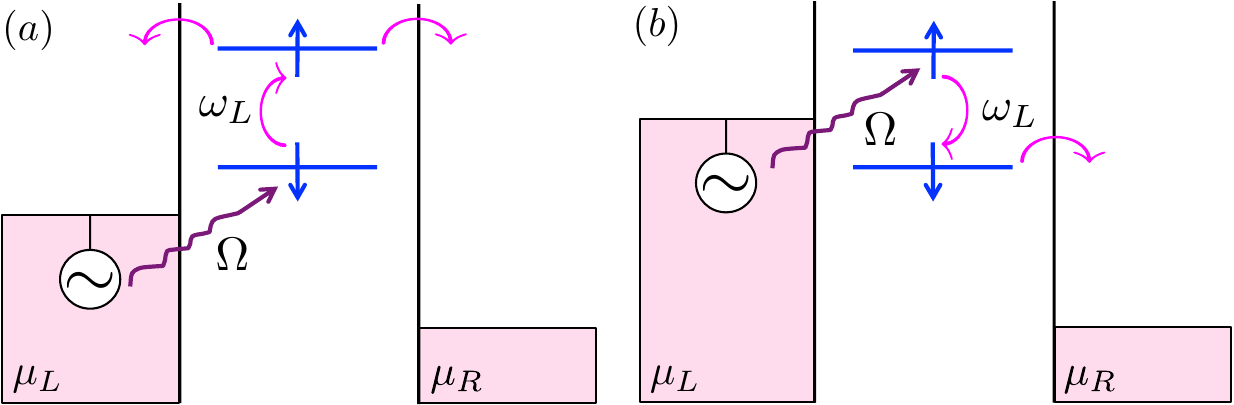}  
	\caption{(Color online) Sketch of two opposite photon-assisted spin-flip processes between molecular quasienergy levels in the presence of ac harmonic potential with frequency $\Omega$ in the left lead. (a) Excited electron with energy $\Omega$ tunnels into spin-down level $\epsilon_{\downarrow}$ (or $\epsilon_{\uparrow}-\omega_L$). It absorbs an amount of energy $\omega_L$, flips its spin due to the exchange interaction with the precessing component of the molecular spin, and exits into either lead. (b) Excited electron tunnels into spin-up level $\epsilon_{\downarrow}+\omega_{L}$ (or $\epsilon_{\uparrow}$), flips its spin, and emits an energy quantum $\omega_L$. Then it tunnels out to the right lead.}\label{fig: sketch}
\end{figure}

Some of the photon-assisted tunneling processes contributing to the spin transport are presented in Fig.~\ref{fig: sketch}. We show examples of two opposite photon-assisted spin-flip processes. 
Fig.~\ref{fig: sketch}(a) corresponds to the case in which $\epsilon_{1}-\mu_{L}<\Omega$ (or $\epsilon_{3}-\mu_{L}<\Omega$). Here an electron from the left lead excited by a photon of energy $\Omega=\omega_L$ tunnels into the level $\epsilon_1$ (or $\epsilon_3$). During the exchange interaction with the precessing component of $\vec{S}(t)$, it
absorbs an energy $\omega_L$ and flips its spin, ending up in the level $\epsilon_{2}$ (or $\epsilon_4$), and then tunnels into either of the leads.
One photon-assisted spin-flip process through level $\epsilon_{2}$ (or $\epsilon_{4}$) for $\epsilon_{1}\leq\mu_{L}\leq\epsilon_{2}$ (or $\epsilon_{3}\leq\mu_{L}\leq\epsilon_{4}$) is presented in Fig.~\ref{fig: sketch}(b). In this case, an electron absorbs an energy $\Omega=\omega_L$ interacting with ac field in the left lead and enter the spin-up level $\epsilon_2$ (or $\epsilon_4$). Then, it emits energy quantum $\omega_{L}$, flips its spin due to the interaction with the precessing component of the molecular spin, and tunnels into the right lead.

\begin{figure}
	\includegraphics[height=4.9cm,keepaspectratio=true]{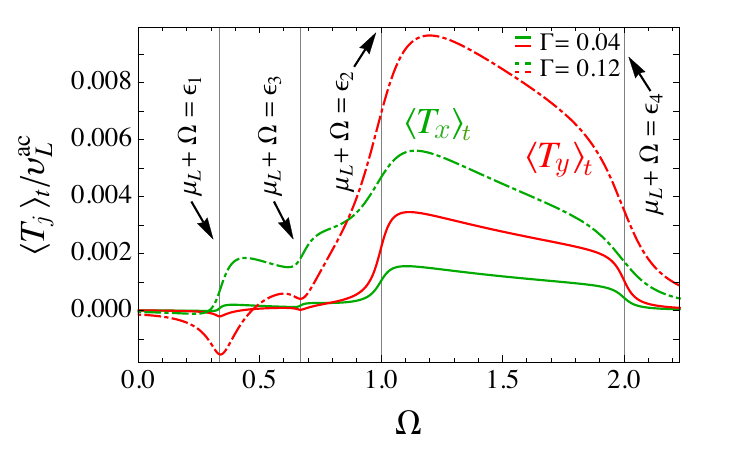}   
	\caption{(Color online) Time-averaged spin-transfer torque components $\langle T_{j}\rangle_{t}$ for $j=x,y$ as functions of ac frequency $\Omega$. The plots are obtained at zero temperature for two different $\Gamma$, with $\Gamma_{L}=\Gamma_{R}=\Gamma/2$, $\vec{B}=B\vec{e}_{z}$, and $\Omega=\omega_{L}$. All energies are given in the units of $\epsilon_0$. The other parameters are set to: $\mu_{L}=0.25, \, \mu_{R}=0, \, \phi_{L}=1.95, \, v^{\rm{ac}}_{R}=0, \,\phi_{R}=0$,
		$\theta=1.25,\allowbreak J=0.005$, and 
		$S=\nobreak100$. Each step or peak coincides with a change in the number of available channels for photon-assisted spin tunneling.}\label{fig: torque k}
\end{figure}
\begin{figure}
	\includegraphics[height=4.9cm,keepaspectratio=true]{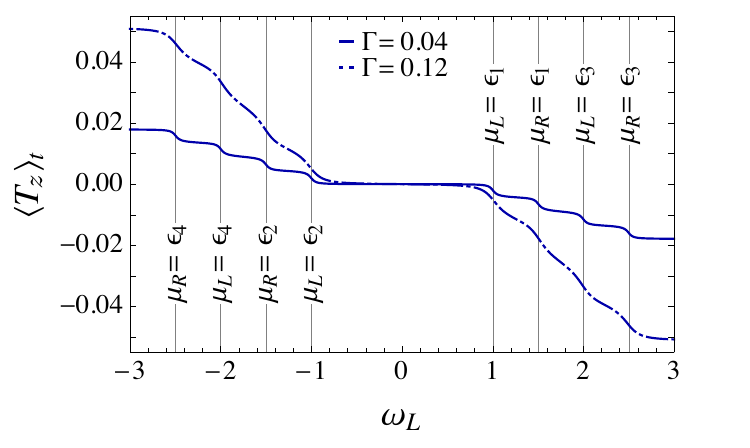}   
	\caption{(Color online) Time-averaged $z$ component of the spin-transfer torque $\langle T_{z}\rangle_t$ as a function of the Larmor precession frequency $\omega_L$. The plots are obtained for two different tunneling rates $\Gamma$ at zero temperature, with $\vec{B}=B\vec{e}_{z}$ and $\Gamma_{L}=\Gamma_{R}=\Gamma/2$. All energies are given in the units of $\epsilon_0$. The other parameters are set to: $\mu_{L}=0.25, \, \mu_{R}=0, \,
		\theta=1.25,\allowbreak J=0.005$, and
		$S=100$. Each step corresponds to a spin-tunneling process involving a spin-flip.}\label{fig: torque z}
\end{figure}

In Fig.~\ref{fig: torque k} the time-averaged $x$ and $y$ components of STT are plotted as functions of ac frequancy $\Omega=\omega_L$, for two different tunnel coupling constants $\Gamma=0.04$ (solid lines) and $\Gamma=0.12$ (dot-dashed lines), at zero temperature. The grid lines correspond to $\epsilon_{i}-\mu_{L}=\Omega$.
For $\Omega$ such that $\epsilon_{1}-\mu_{L}=\Omega$ the molecular quasienergy level $\epsilon_1$ participates in photon-assisted spin transport, followed by an electron spin-flip and hence a finite STT. In this case $\langle T_{x}\rangle_t$ is initially enhanced while $\langle T_{y}\rangle_t$ has a minimum value and increases after $\Omega=\epsilon_{1}-\mu_{L}$ [first grid line in Fig. \ref{fig: torque k}]. As $\Omega$ increases the inequality (\ref{eq:inequal}) is satisfied for level $\epsilon_1$ leading to a nonzero STT. With further increase of ac frequency $\Omega$ the photon-assisted spin transport begins to take place in the level $\epsilon_3$. Both $\langle T_{x}\rangle_t$ and $\langle T_{y}\rangle_t$ increase around $\Omega=\epsilon_{3}-\mu_{L}$, after going to a local minimum, due to the fact that level $\epsilon_3$ is now available for spin-flip tunneling processes.

For larger $\Omega$ the inequality (\ref{eq:inequal}) is satisfied for both $\epsilon_1$ and $\epsilon_3$. Consequently, both $\langle T_{x}\rangle_t$, and $\langle T_{y}\rangle_t$ increase. Finally, as $\Omega$ increases further, level $\epsilon_2$ also becomes available for photon-assisted spin tunneling, leading to the largest enhancement of both in-plane STT components. As $\Omega$ increases further, inequality (\ref{eq:inequal}) is satisfied for levels $\epsilon_1$, $\epsilon_2$, and $\epsilon_3$, and photon-assisted STT components are large and decreasing. After the level $\epsilon_4$ becomes available for photon-assisted spin transport, both components $\langle T_{x}\rangle_t$ and $\langle T_{y}\rangle_t$ drop to zero. This is due to the previously mentioned antisymmetry. Namely, in this case, the contributions of the photon-assisted STTs for $\epsilon_{1}<\mu_{L}<\epsilon_{2}$ and $\epsilon_{3}<\mu_{L}<\epsilon_{4}$ are equal in magnitude but have opposite directions. Therefore, they cancel each other as $\mu_L$ satisfies both these inequalities simultaneously. Conditions of inequality (\ref{eq:inequal}) are relaxed for larger $\Gamma$ due to the broadening of the levels $\epsilon_i$.

The $z$ component of the time-averaged STT, $\langle T_{z}\rangle_t$ is plotted as a function of the Larmor frequency $\omega_L$ in Fig.~\ref{fig: torque z}. This component does not contain contributions from photon-assisted spin tunneling, but only from tunneling under dc-bias voltage, followed by an electron spin flip due to the interaction with the precessing component of the molecular spin $\vec{S}(t)$. In turn, an STT is exerted on the molecular spin. The STT component $\langle T_{z}\rangle_t$ is an odd function of $\omega_L$ since the change of the direction of $\vec{B}$ gives negative $\omega_L$.
Each step in Fig.~\ref{fig: torque z} denotes a new available spin-transport channel, and an additional spin-flip process, contributing to the STT, which takes place for $\mu_{\xi}=\epsilon_i$.

\section{Conclusions}

In this paper we have theoretically studied photon-assisted spin and charge transport through a molecular magnet junction. The junction consists of single molecular orbital in the presence of a molecular spin precessing with Larmor frequency $\omega_L$ in a constant magnetic field. The orbital is connected to two metal leads subject to harmonically varying chemical potentials with frequency $\Omega$, treated as a perturbation. We used the Keldysh nonequilibrium Green's functions method to derive charge and spin currents and the spin-transfer torque. We employed the displacement current partitioning scheme of Wang \textit{et al.}\cite{Guo1} to obtain gauge invariant expressions for the dynamic conductance of the charge current.
The dynamic response of the system is controlled by photon-assisted transport. In the low-ac-frequency limit this junction displays an inductive-like or capacitive-like behavior depending on the system parameters. When the chemical potentials are in resonance with a molecular quasienergy level $\epsilon_i$, the real and imaginary components of the ac conductance both increase around the ac frequency which coincides with the Larmor frequency, after going to a local minimum, thus allowing to reveal the Larmor frequency by a conductance measurement.
The photon-assisted $x$ and $y$ spin-current components consist of a dc part and a part that oscillates with frequency $2\omega_L$ for $\Omega=\omega_L$. This opens a possibility to experimentally investigate photon-assisted spin-transfer torque exerted on the molecular magnet, which can be detected through the presence of  nonzero time-averaged contributions.
By manipulating the phases of the harmonic potentials in the leads with respect to the precession, and the tilt angle between the magnetic field and the molecular spin, the control of the direction and the magnitude of the time-averaged photon-assisted spin current components and spin-transfer torque is achievable. Finally, in this work we present the nonperturbative Gilbert damping and other STT coefficients with respect to the coupling $\gamma$, in the zero ac frequency limit. Remarkably, the Gilbert damping vanishes in the strong-coupling limit.

In the future it might be interesting to investigate further transport properties like the current noise or the spin-torque noise, as well as to find ways to manipulate the magnetic moment using e.g., ferromagnetic leads.

\begin{acknowledgments}
	We would like to thank Gianluca Rastelli, Federica Haupt and Cecilia Holmqvist for useful discussions. We gratefully acknowledge the financial support from the Deutsche Forschungsgemeinschaft through the SFB 767 \textit{Controlled Nanosystems} and the SPP 1285 \textit{Semiconductor Spintronics} and an ERC Advanced Grant \textit{UltraPhase} of Alfred Leitenstorfer.
\end{acknowledgments}

\begin{widetext}

\appendix*

\section*{Appendix: Expressions for spin-current components}
     \renewcommand{\theequation}{A\arabic{equation}}
       \setcounter{equation}{0}

Here we present the expressions for spin-current complex components $I_{Lx}(\omega_L)$ and $I_{Ly}(\omega_L)$, spin current $I^{\omega_L}_{Lz}$ in the presence of dc-bias volatge, and spin currents in the presence of ac voltage in terms of the matrix elements of the Green's functions $\hat{G}^{0r}(\epsilon)$ and $\hat{G}^{0a}(\epsilon)$.

The expressions for spin-current complex components introduced by Eq.~(\ref{eq: k resonant}) are given by

	\begin{align}
	I_{Lx}(\omega_L)=\label{eq: x resonant}&-i\int\frac{d\epsilon}{4\pi}\Bigg\{\frac{\Gamma_{L}\Gamma_{R}}{\Gamma}[f_{L}(\epsilon)-f_{R}(\epsilon)]
	\Bigg[\frac{\gamma G^{0r}_{11}(\epsilon+\omega_{L})G^{0r}_{22}(\epsilon)}
	{\lvert 1-\gamma^{2}G^{0r}_{11}(\epsilon+\omega_{L})G^{0r}_{22}(\epsilon)\lvert^{2}}\nonumber\\
	&+\frac{2i\gamma{\rm Im}
		\{G^{0r}_{11}(\epsilon)\}G^{0a}_{22}(\epsilon-\omega_{L})+\gamma^{3}\lvert G^{0r}_{11}(\epsilon) G^{0r}_{22}(\epsilon-\omega_{L})\lvert^{2}}
	{\lvert 1-\gamma^{2}G^{0r}_{11}
		(\epsilon)G^{0r}_{22}(\epsilon-\omega_{L})\lvert^{2}}\Bigg]\nonumber\\
	&+\sum_{\xi,\zeta=L,R}\frac{\Gamma_{\xi}\Gamma_{\zeta}}{\Gamma}[f_{\xi}(\epsilon-\omega_{L})-f_{\zeta}(\epsilon)]
	\Big [\delta_{\zeta L}-\delta_{\xi L}\gamma^{2}G^{0a}_{11}(\epsilon)G^{0r}_{22}(\epsilon-\omega_{L})\Big ]
	\frac{\gamma G^{0r}_{11}(\epsilon)G^{0a}_{22}(\epsilon-\omega_{L})}
	{\lvert 1-\gamma^{2}G^{0r}_{11}(\epsilon)G^{0r}_{22}(\epsilon-\omega_{L})\lvert^{2}}\Bigg \},\\
	I_{Ly}(\omega_L)=\label{eq: y resonant}&\,\ iI_{Lx}(\omega_L),\\ 
	{\rm while}\,\,\ I_{Lz}^{\omega_{L}}=\label{eq: z resonant}&\int\frac{d\epsilon}{4\pi}\Bigg\{\frac{\Gamma_{L}\Gamma_{R}}{\Gamma}[f_{L}(\epsilon)-f_{R}(\epsilon)]
	\Bigg[\frac{2{\rm Im}\{G^{0r}_{11}(\epsilon)\}}
	{\lvert 1-\gamma^{2}G^{0r}_{11}(\epsilon)G^{0r}_{22}(\epsilon-\omega_{L})\lvert^{2}}-\frac{2{\rm Im}\{G^{0r}_{22}(\epsilon)\}}
	{\lvert 1-\gamma^{2}G^{0r}_{11}(\epsilon+\omega_{L})G^{0r}_{22}(\epsilon)\lvert^{2}}\Bigg]\nonumber\\
	&+\sum_{\xi,\zeta=L,R}\Gamma_{\xi}\Gamma_{\zeta}[f_{\xi}(\epsilon-\omega_{L})-f_{\zeta}(\epsilon)]
	(\delta_{\xi L}+\delta_{\zeta L})
	\frac{\gamma^{2}\lvert G^{0r}_{11}(\epsilon) G^{0r}_{22}(\epsilon-\omega_{L})\lvert^{2}}
	{\lvert 1-\gamma^{2}G^{0r}_{11}(\epsilon)G^{0r}_{22}(\epsilon-\omega_{L})\lvert^{2}}\Bigg \}.
	\end{align}
The spin-current components in the presence of oscillating chemical potentials in the leads, introduced as the second term in Eq.~(\ref{eq:particle_current}), for $\xi=L$ can be expressed in the following way:
	\begin{align}
		I^{\Omega}_{Lj}(t)=\label{eq: k assisted}&\sum_{\xi=L,R}{\rm Re}\Big\{[I^{j}_{L\xi}(\Omega)e^{-i(\Omega t+\phi_{\xi})}+I^{j}_{L\xi}(-\Omega)e^{i(\Omega t+\phi_{\xi})}]
		e^{-i\omega_{L}t}\Big\},\qquad 
		\qquad j=x,y,\\
		{\rm where} \quad I^{x}_{L\xi}(\Omega)=\gamma\label{eq: x assisted}&\Gamma_{L}\Gamma_\xi\frac{ v_{\xi}^\mathrm{ac}}{\Omega}\int\frac{d\epsilon}{4\pi}[f_{\xi}(\epsilon-\Omega)-f_{\xi}(\epsilon)]\nonumber\\
		\times &\Bigg\{\frac{G^{0a}_{11}(\epsilon-\Omega)G^{0a}_{22}(\epsilon-\Omega-\omega_L)
			\{G^{0r}_{11}(\epsilon)+i\frac{\delta_{L\xi}}{\Gamma_\xi}[1-\gamma^{2}G^{0r}_{11}(\epsilon)
			G^{0r}_{22}(\epsilon-\omega_{L})]\}}
		{[1-\gamma^{2}G^{0r}_{11}(\epsilon)G^{0r}_{22}(\epsilon-\omega_{L})][1-\gamma^{2}G^{0a}_{11}(\epsilon-\Omega)G^{0a}_{22}(\epsilon-\Omega-\omega_{L})]}\nonumber\\
		&+\frac{G^{0r}_{11}(\epsilon+\omega_L)G^{0r}_{22}(\epsilon)
			\{G^{0a}_{22}(\epsilon-\Omega)-i\frac{\delta_{L\xi}}{\Gamma_\xi}[1-\gamma^{2}
			G^{0a}_{11}(\epsilon-\Omega +\omega_{L})G^{0a}_{22}(\epsilon-\Omega)]\}}
		{[1-\gamma^{2}G^{0a}_{11}(\epsilon-\Omega+\omega_{L})G^{0a}_{22}(\epsilon-\Omega)][1-\gamma^{2}G^{0r}_{11}(\epsilon+\omega_{L})G^{0r}_{22}(\epsilon)]}\Bigg\},\\
		{\rm while}\quad I^{y}_{L\xi}(\Omega)=\label{eq: y assisted}&\,\ i I^{x}_{L\xi}(\Omega),\\
		{\rm and} \quad I^{\Omega}_{L z}(t)=&\sum_{\xi=L,R}\sum_{\substack{\sigma=\pm 1}}{\rm Re}\Bigg\{\Gamma_{L}\Gamma_\xi\frac{v_{\xi}^\mathrm{ac}}{\Omega}\int\frac{d\epsilon}{4\pi}
		[f_{\xi}(\epsilon-\Omega)-f_{\xi}(\epsilon)]e^{-i(\Omega t+\phi_{\xi})}\label{eq: z assisted}\nonumber\\
		&\times\frac{[\hat{\sigma}_{z}\hat{G}^{0r}(\epsilon)\hat{G}^{0a}(\epsilon-\Omega)]_{\sigma\sigma}
			\Big\{2-[1-i\frac{\delta_{L\xi}}{\Gamma_\xi}(\Omega+i\Gamma)][1+\gamma^{2}
			G^{0r}_{-\sigma-\sigma}(\epsilon_{\sigma})
			G^{0a}_{-\sigma-\sigma}(\epsilon_{\sigma}-\Omega)]\Big\}}
		{[1-\gamma^{2}G^{0r}_{\sigma\sigma}(\epsilon)G^{0r}_{-\sigma-\sigma}(\epsilon_{\sigma})][1-\gamma^{2}G^{0a}_{\sigma\sigma}
			(\epsilon-\Omega)G^{0a}_{-\sigma-\sigma}(\epsilon_{\sigma}-\Omega)]}\Bigg\}.
	\end{align} 
\end{widetext}

\end{document}